\documentclass[aps,pra,twocolumn,showpacs,superscriptaddress]{revtex4}
\usepackage{graphicx}
\usepackage{amsmath}
\usepackage{amssymb}
\usepackage{lscape}
\usepackage{epstopdf}

\input{epsf}
\begin{document}

\title{Harmonically trapped two-atom systems:
Interplay of short-range $s$-wave interaction
and spin-orbit coupling}

\author{X. Y. Yin}
\affiliation{Department of Physics and Astronomy,
Washington State University,
  Pullman, Washington 99164-2814, USA}
\author{S. Gopalakrishnan}
\affiliation{Department of Physics, Harvard University,
Cambridge, Massachusetts 02138, USA}
\author{D. Blume}
\affiliation{Department of Physics and Astronomy,
Washington State University,
  Pullman, Washington 99164-2814, USA}

\date{\today}

\pacs{03.75.Mn, 05.30.Fk, 05.30.Jp, 67.85.Fg}

\begin{abstract}
The coupling between the spin degrees of freedom and the orbital angular
momentum has a profound effect on 
the properties of nuclei,
atoms and condensed matter systems. 
Recently, synthetic gauge fields have been realized experimentally in 
neutral cold atom systems,
giving rise to a spin-orbit coupling 
term with ``strength'' $k_{\text{so}}$.
This paper investigates the interplay between the single-particle
spin-orbit coupling term of Rashba type and the short-range
two-body $s$-wave interaction for cold 
atoms under external  confinement.
Specifically, we consider two different harmonically trapped two-atom
systems.
The first system consists of an atom with spin-orbit coupling that
interacts with a structureless particle through 
a short-range two-body potential.
The second system consists of two atoms that
both feel the spin-orbit coupling term and that
interact  through a short-range two-body
potential.
Treating the spin-orbit term perturbatively, we
determine the correction to the ground state energy 
for various generic parameter combinations.
Selected excited states are also treated.
An important aspect of our study is that the perturbative treatment
is not limited to small $s$-wave scattering lengths but
provides insights into the system behavior over a wide range of scattering
lengths, including the strongly-interacting unitary regime.
We find that the interplay between the spin-orbit
coupling term and the $s$-wave interaction generically enters,
depending on the exact parameter combinations of the
$s$-wave scattering lengths, at order $k_{\text{so}}^2$
or $k_{\text{so}}^4$ for the ground state and leads to a
shift of the energy of either sign.
While the absence of a term proportional to
$k_{\text{so}}$ follows straightforwardly from the functional form
of the spin-orbit coupling term, the absence of a term
proportional to  $k_{\text{so}}^2$ for certain parameter 
combinations is unexpected.
The well-known fact that the spin-orbit coupling 
term couples the relative and center of mass degrees
of freedom has interesting consequences
for the trapped two-particle systems.
For example, we find that the spin-orbit
coupling term turns, for certain parameter combinations,
sharp crossings into avoided crossings
with an energy splitting proportional to $k_{\text{so}}$.
Our perturbative results are confirmed by numerical calculations
that expand the eigenfunctions of the two-particle
Hamiltonian  in terms of basis
functions that contain explicitly correlated Gaussians.
\end{abstract}

\maketitle

\section{Introduction}
During the past few years
tremendous progress has been made in realizing artificial gauge fields
in cold atom systems experimentally~\cite{rmp11, 
spielmannat13, spielmanreview13, zhaireview}.
By now, the effect of the spin-orbit coupling (or more precisely, spin-momentum
coupling) has been investigated for bosonic and fermionic
species~\cite{spielman11, jzhang12, spielman13, jzhang13,
engels13, greene13, zwierlein12}.
The effect of the spin-orbit coupling 
has been investigated away and near an $s$-wave 
Fano-Feshbach resonance~\cite{spielman13, jzhang13}.
A variety of intriguing phenomena such as non-equilibrium
dynamics~\cite{engels13, greene13}, the spin-orbit coupling
assisted formation of molecules~\cite{jzhang13}, and the engineering
of band structures~\cite{zwierlein12} have been investigated.

At the mean-field level, spin-orbit coupled gases exhibit rich phase diagrams~\cite{zhaireview, wu11, zhai10, pu11, pu12, zhai11, 
galitski08, czhang11}.
Effects beyond mean-field 
theory~\cite{sarang11, baym12, baym11, williams11}, 
associated with the renormalization of interactions, 
are enhanced by the spin-orbit coupling, 
especially in the pure Rashba case, 
and can qualitatively change the mean-field results. 
Thus, the interplay between the spin-orbit coupling 
and the $s$-wave interaction is a 
crucial aspect of the many-body physics of such systems.
The two-particle scattering for systems with spin-orbit
coupling has been investigated using a variety of different 
approaches~\cite{pzhang13, gao13, cui12, pzhang12a,
pzhang12b}
including a Green's function approach and a 
quantum defect theory approach.
Compared to the scattering between two alkali
atoms, the scattering between particles with spin-orbit
coupling introduces a coupling between different partial wave channels.
Moreover, if the two-particle system
with Rashba spin-orbit coupling is loaded into an
external harmonic trap, the relative and the center of mass degrees
of freedom do not decouple.

This paper determines the quantum mechanical energy spectrum of two
atoms with short-range two-body interactions
in an external spherically symmetric
harmonic trap in the presence of a Rashba spin-orbit coupling term.
Our work combines analytical and numerical approaches,
and covers weak spin-orbit coupling strengths
and weak to strong atom-atom interactions.
Few atom systems can nowadays
be prepared and probed experimentally~\cite{jochim11, bloch10}, 
opening the door
for developing a bottom-up understanding of cold
atom systems with spin-orbit coupling.
Our results provide much needed theoretical guidance for
such experimental studies.
Two prototype systems of increasing complexity
are considered.
{\em{(i)}} We assume that one of the particles 
feels the Rashba coupling while the other
does not.
{\em{(ii)}} We assume that both particles feel the Rashba
coupling.
The first system under study can also be viewed as the limiting
case of a two-component atomic gas where
one component feels the spin-orbit coupling term while
the other does not. While such systems have not
yet been realized experimentally, their
preparation is feasible with current technology.
The second system under study can be viewed as a limiting 
case of a bosonic or fermionic gas with spin-orbit coupling.
Our analysis of the two-particle prototype systems yields, e.g., 
an analytical expression for the leading-order mean-field shift
that reflects the interplay between the spin-orbit coupling term
and the $s$-wave interaction.

The effect of spin-orbit coupling has also been studied
in condensed matter systems, such as
two-dimensional electron gases~\cite{2degexp, 2degtheory}, 
semiconductor quantum dots~\cite{qdexp, governale02,
pietilainen05, pi09, reimann11} and 
semiconductor nanowires~\cite{nanoexp}. 
Employing a perturbative expansion for the
two-dimensional electron gas, the long-range 
electron-electron interactions have been found to
be influenced only marginally by the spin-orbit
coupling~\cite{2degtheory},
in qualitative agreement with our findings for short-range
$s$-wave interactions. 
Just as the atoms considered in this work,
the electrons in semiconductor quantum dots are subject to
a confining potential that is well approximated by a harmonic
trap and feel a Rashba spin-orbit coupling term. 
In many materials the Rashba term, which is tunable to some extent,
dominates over the Dresselhaus term.
Much attention has been paid to the interplay between the electron-electron
interaction and the spin-orbit coupling term~\cite{governale02,
pietilainen05, reimann11}.
While similar in spirit, key differences between the quantum dot
studies and our work exist: {\em{(i)}} The electron-electron interaction 
is long-ranged and repulsive while the atom-atom interaction 
considered in this work is short-ranged and effectively repulsive
or effectively attractive. 
 {\em{(ii)}} Electrons obey fermionic
statistics while our work considers fermionic and bosonic atoms.
 {\em{(iii)}} The quantum dots are typically modeled
assuming a two-dimensional confining geometry while our work considers a three-dimensional confining geometry.

The remainder of this paper is organized as follows.
Section~\ref{sec_systemham} defines the system Hamiltonian.
Section~\ref{sec_weakall} investigates the regime where the
spin-orbit coupling strength and the atom-atom interaction are
weak. A perturbative approach that yields analytic energy
expressions is developed. As we will show, this approach provides
valuable insights into the interplay of the spin-orbit
coupling term and the atom-atom interaction.
Section~\ref{sec_weaknotall} develops a complementary
perturbative approach. Namely, accounting for the atom-atom
interaction exactly~\cite{busc98},
the spin-orbit coupling term is treated as a perturbation. This approach
provides valuable insights into the system dynamics over a wide
range of scattering lengths, including the unitary regime.
Our perturbative results of Secs. III and IV are validated by numerical results. The discussion of the numerical 
approach that yields accurate eigenenergies of the 
trapped two-particle system
is relegated to the Appendix.
Section~\ref{sec_conclusion} summarizes and offers an outlook.

\section{System Hamiltonian}
\label{sec_systemham}
We consider two particles 
of mass $m$ with position vectors 
$\vec{r}_{j}=(x_{j},y_{j},z_{j})$, where $j=1$ and 2.
The position vectors
are measured with respect to the center of the harmonic trap
(see below) 
and 
the distance vector is denoted by $\vec{r}_{12}$,
$\vec{r}_{12}=\vec{r}_1-\vec{r}_2$ and
$r_{12}=|\vec{r}_{12}|$.
This paper considers two different situations:
In the first case, 
the first atom feels the spin-orbit coupling of Rashba type
while the second atom does not.
In the second case, both atoms feel the spin-orbit coupling
of Rashba type.
If the $j$th atom feels the spin-orbit coupling, it is assumed to have
two internal states
denoted by $| \uparrow \rangle_j$
and
$| \downarrow \rangle_j$.
As commonly done,
we identify the two internal states of the $j$th
atom as pseudo-spin states 
of a spin-1/2 particle with 
spin projection quantum numbers $m_{sj}=1/2$ and $m_{sj}=-1/2$.
Concretely, the spin-orbit coupling term $V_{\text{so}}(\vec{r}_j)$ of
the $j$th atom reads~\cite{rashba84}
\begin{eqnarray}
\label{eq_vso}
V_{\text{so}}(\vec{r}_{j})=
-\imath \frac{\hbar^2 k_{\text{so}}}{m}
\Bigg[
\left(
\frac{\partial}{\partial y_{j}} + \imath \frac{\partial}{\partial x_{j}}
\right)
| \uparrow \rangle_j \, _j\langle \downarrow |
+ \nonumber \\
\left(
\frac{\partial}{\partial y_{j}} - \imath \frac{\partial}{\partial x_{j}}
\right)
| \downarrow \rangle_j \, _j\langle \uparrow |
\Bigg].
\end{eqnarray}

If only the first particle feels the spin-orbit coupling,
the Hamiltonian $H_{\text{soc,a}}$  of the harmonically trapped 
two-particle system can be written as
\begin{eqnarray}
\label{eq_ham_case1}
H_{\text{soc,a}}= H^{(1)}(\vec{r}_1) + H_{\text{ho}}(\vec{r}_{2}) + 
H^{(12)}_{\text{soc,a}}(\vec{r}_{12}).
\end{eqnarray}
If both atoms feel the spin-orbit coupling,
the Hamiltonian $H_{\text{soc,soc}}$  of the harmonically trapped 
two-particle system can be written as
\begin{eqnarray}
\label{eq_ham_case2}
H_{\text{soc,soc}}= H^{(1)}(\vec{r}_1) + H^{(1)}(\vec{r}_{2}) + 
H_{\text{soc,soc}}^{(12)}(\vec{r}_{12}).
\end{eqnarray}
In Eqs.~(\ref{eq_ham_case1}) and (\ref{eq_ham_case2}), $H^{(1)}$
denotes the single-atom Hamiltonian,
\begin{eqnarray}
\label{eq_ham_a}
H^{(1)}(\vec{r}_j) = 
\sum_{\sigma= \uparrow,\downarrow}
H_{\text{ho}}(\vec{r}_{j})
|\sigma \rangle_j \, _j \langle \sigma| + 
V_{\text{so}}(\vec{r}_j),
\end{eqnarray}
and
$H_{\text{ho}}(\vec{r}_j)$ the three-dimensional single-particle harmonic
oscillator Hamiltonian with angular frequencies $\omega_x$, $\omega_y$ and
$\omega_z$,
\begin{eqnarray}
\label{eq_ham_ho}
H_{\text{ho}}(\vec{r}_j)=-\frac{\hbar^2}{2m} 
\left(
\frac{\partial^2}{\partial x_j^2}+
\frac{\partial^2}{\partial y_j^2}+
\frac{\partial^2}{\partial z_j^2} \right)
+ \nonumber \\
\frac{1}{2} m (\omega_x^2 x_j^2+ \omega_y^2 y_j^2+ \omega_z^2 z_j^2).
\end{eqnarray}
Throughout most of this paper, we assume $\omega_x=\omega_y=\omega_z=\omega$.
Correspondingly, we measure lengths in units of 
$a_{\text{ho}}$, where $a_{\text{ho}}=\sqrt{\hbar/(m \omega)}$,
and energies in units of $E_{\text{ho}}$, where $E_{\text{ho}}=\hbar \omega$.
We note, however, that the techniques
developed in this work can be generalized to
anisotropic confinement.
In Eqs.~(\ref{eq_ham_case1}) and (\ref{eq_ham_case2}),
$H^{(12)}_{\text{soc,a}}(\vec{r}_{12})$
and
$H^{(12)}_{\text{soc,soc}}(\vec{r}_{12})$
account for the atom-atom interaction.
We note that the single particle Hamiltonian $H^{(1)}(\vec{r}_j)$
and variants thereof have been investigated extensively in quantum optics 
and molecular physics~\cite{doucha87, analytic}. 
In quantum optics the Hamiltonian
is referred to as the Jaynes-Cummings Hamiltonian.  
In molecular physics, the Hamiltonian is referred to as the
$E \otimes \epsilon$ Jahn-Teller Hamiltonian.

If both particles feel the spin-orbit coupling,
we assume an interaction of the form
\begin{eqnarray}
\label{eq_ham_socsoc}
H^{(12)}_{\text{soc,soc}}(\vec{r}_{12})
=
V_{\text{2b}}^{\uparrow \uparrow}(\vec{r}_{12}) 
|\uparrow \rangle_1 | \uparrow \rangle_2 \, 
_1 \langle \uparrow | _2 \langle \uparrow |
+  \nonumber \\
V_{\text{2b}}^{\uparrow \downarrow}(\vec{r}_{12}) 
|\uparrow \rangle_1 | \downarrow \rangle_2 \, 
_1 \langle \uparrow | _2 \langle \downarrow |
+ \nonumber \\
V_{\text{2b}}^{\downarrow \uparrow}(\vec{r}_{12}) 
|\downarrow \rangle_1 | \uparrow \rangle_2 \, 
_1 \langle \downarrow | _2 \langle \uparrow |
+ \nonumber \\
V_{\text{2b}}^{\downarrow \downarrow}(\vec{r}_{12}) 
|\downarrow \rangle_1 | \downarrow \rangle_2 \, 
_1 \langle \downarrow | _2 \langle \downarrow |
.
\end{eqnarray}
The potentials $V_{\text{2b}}^{\sigma \sigma'}(\vec{r}_{12})$
($\sigma,\sigma'=\uparrow$ or $\downarrow$) 
are characterized by the scattering lengths $a_{\sigma \sigma'}$.
We write $a_{\uparrow \uparrow}=a_{\text{aa}}$,
$a_{\downarrow \downarrow}=\zeta a_{\text{aa}}$
and $a_{\uparrow \downarrow}=a_{\downarrow \uparrow}=\eta a_{\text{aa}}$.
Experimentally, the scattering lengths can,
in certain cases, be tuned by applying an external magnetic field in 
the vicinity of a Fano-Feshbach resonance~\cite{chinrmp}. 
We consider three different interaction models, 
a zero-range $s$-wave pseudo-potential
$V_{\text{ps}}^{\sigma \sigma'}(\vec{r}_{12})$ 
with 
scattering length $a_{\sigma \sigma'}$, a regularized pseudo-potential
$V_{\text{ps,reg}}^{\sigma \sigma'}(\vec{r}_{12})$, 
and a Gaussian model potential $V_{\text{g}}^{\sigma \sigma'}(\vec{r}_{12})$ 
with range $r_{0}$ and 
depth/height $V_{0}^{\sigma \sigma'}$,
\begin{eqnarray}
\label{eq_pot_pseudo}
V_{\text{ps}}^{\sigma \sigma'}(\vec{r}_{12})= 
\frac{4 \pi \hbar^2 a_{\sigma \sigma'}}{m} \delta(\vec{r}_{12}),
\end{eqnarray}
\begin{eqnarray}
\label{eq_pot_pseudoreg}
V_{\text{ps,reg}}^{\sigma \sigma'}(\vec{r}_{12})= 
\frac{4 \pi \hbar^2 a_{\sigma \sigma'}}{m} \delta(\vec{r}_{12})
\frac{\partial}{\partial r_{12}} r_{12},
\end{eqnarray}
and
\begin{eqnarray}
\label{eq_pot_gaussian}
V_{\text{g}}^{\sigma \sigma'}(\vec{r}_{12}) = V_{0}^{\sigma \sigma'} \exp \left[ -
\left( \frac{r_{12}}{\sqrt{2} r_{0}} \right)^2 \right].
\end{eqnarray}
To compare the results for the
zero-range and finite-range potentials, the parameters
$r_0$ and 
$V_0^{\sigma \sigma'}$ are adjusted so as to produce the desired 
free-space atom-atom $s$-wave scattering lengths
$a_{\sigma \sigma'}$. We work in the parameter
space where $V_{\text{g}}^{\sigma \sigma'}$ supports either no or one 
free-space $s$-wave bound state.

To date, spin-orbit coupling terms (although not of Rashba type)
have been realized using $^{87}$Rb, $^{7}$Li and $^{40}$K.
In $^{87}$Rb, 
the spin-up and spin-down states are commonly
identified with the $|F, M_F \rangle=|1, 0\rangle$ 
and $|1, -1\rangle$ states~\cite{spielman11, engels13, greene13}. 
The corresponding scattering lengths are 
$a_{\uparrow \uparrow}=100.86a_0$,
$a_{\downarrow \downarrow}=100.40a_0$ 
and $a_{\uparrow \downarrow}=100.41a_0$,
where $a_0$ is the Bohr radius~\cite{hamnerthesis}
(implying $\zeta=0.9954$ and $\eta=0.9955$),
and Feshbach resonances do not
exist. For $^{40}$K in the 
$|F, M_F \rangle=|9/2, 9/2\rangle$ 
and $|9/2, 7/2\rangle$ states~\cite{jzhang12}
or $|F, M_F \rangle=|9/2, -7/2\rangle$ 
and $|9/2, -9/2\rangle$ states~\cite{jzhang13, spielman13},
in contrast,
the $a_{\uparrow \downarrow}$ scattering length
is tunable while $s$-wave scattering is forbidden for the
up-up and down-down channels.
The present work considers cases 2a-2d
(see Table~\ref{table_cases}).
The parameter combination
$a_{\downarrow \downarrow} = a_{\uparrow \downarrow} =
a_{\downarrow \uparrow} \neq a_{\uparrow \uparrow}$
is equivalent to case 2c if we switch the role of 
$a_{\uparrow \uparrow}$ and $a_{\downarrow \downarrow}$.

\begin{table}
\caption{Summary of the different scattering length
combinations investigated in this work for
one particle with and one without spin-orbit coupling
(described by $H_{\text{soc,a}}$) and for both
particles with spin-orbit coupling
(described by $H_{\text{soc,soc}}$).
Throughout, we write
$a_{\uparrow}=a_{\text{aa}}$ 
and $a_{\downarrow}=\eta a_{\text{aa}}$,
and $a_{\uparrow \uparrow}=a_{\text{aa}}$,
$a_{\downarrow \downarrow}=\zeta a_{\text{aa}}$
and $a_{\uparrow \downarrow}=a_{\downarrow \uparrow}=\eta a_{\text{aa}}$.
}
\label{table_cases}
\centering
\begin{tabular}{c|c|c}
\hline
\hline
$H_{\text{soc,a}}$  &  case 1a &
$a_{\uparrow}\neq a_{\downarrow}$; $\eta\neq1$\\
& case 1b & $a_{\uparrow}=a_{\downarrow}$; $\eta=1$\\
\hline
$H_{\text{soc,soc}}$ &  case 2a &
$a_{\uparrow \uparrow}=a_{\downarrow \downarrow}=
a_{\uparrow \downarrow}=a_{\downarrow \uparrow}$;
$\zeta=1$, $\eta=1$ \\
& case 2b &
$a_{\uparrow \uparrow}=a_{\downarrow \downarrow}\neq
a_{\uparrow \downarrow}=a_{\downarrow \uparrow}$;
$\zeta=1$, $\eta\neq1$ \\
& case 2c &
$a_{\uparrow \uparrow}=a_{\uparrow \downarrow}=
a_{\downarrow \uparrow} \neq a_{\downarrow \downarrow}$;
$\zeta\neq1$, $\eta=1$ \\
& case 2d &
$a_{\uparrow \uparrow} \neq a_{\downarrow \downarrow} \neq
a_{\uparrow \downarrow}=a_{\downarrow \uparrow}$;
$\zeta\neq1$, $\eta\neq1$, $\zeta\neq\eta$ \\
\hline
\hline
\end{tabular}
\end{table} 

If only the first particle feels the spin-orbit coupling,
we assume an atom-atom
interaction of the form
\begin{eqnarray}
\label{eq_ham_soci}
H^{(12)}_{\text{soc,a}}(\vec{r}_{12})
=
V_{\text{2b}}^{\uparrow}(\vec{r}_{12}) 
|\uparrow \rangle_1 \,
_1 \langle \uparrow | 
+  
V_{\text{2b}}^{\downarrow}(\vec{r}_{12}) 
|\downarrow \rangle_1 \, 
_1 \langle \downarrow |.
\end{eqnarray}
The potentials $V_{\text{2b}}^{\uparrow}(\vec{r}_{12})$
and
$V_{\text{2b}}^{\downarrow}(\vec{r}_{12})$
are characterized by the $s$-wave scattering lengths $a_{\uparrow}$
and $a_{\downarrow}$, respectively.
We define $a_{\uparrow}=a_{\text{aa}}$ 
and $a_{\downarrow}=\eta a_{\text{aa}}$,
and consider $\eta=1$  (case 1a) 
and $\eta\neq1$ (case 1b).
As in the case where both particles feel the spin-orbit coupling,
we consider the
zero-range $s$-wave pseudo-potential
$V_{\text{ps}}^{\sigma}(\vec{r}_{12})$, 
the regularized pseudo-potential
$V_{\text{ps,reg}}^{\sigma}(\vec{r}_{12})$, 
and the Gaussian model potential $V_{\text{g}}^{\sigma}(\vec{r}_{12})$.
The definitions of these potentials
are given in Eqs.~(\ref{eq_pot_pseudo})-(\ref{eq_pot_gaussian})
with $\sigma \sigma'$ replaced by 
$\sigma$.

The system Hamiltonian $H_{\text{soc,a}}$ and $H_{\text{soc,soc}}$
are characterized by a number of
length scales:
the harmonic oscillator length $a_{\text{ho}}$, 
the spin-orbit coupling length $1/k_{\text{so}}$, 
and
the atom-atom scattering lengths.
The Gaussian model potential introduces an
additional
length scale, namely the range $r_0$. Throughout this paper,
we consider the regime where $r_{0}$ is
much smaller than $a_{\text{ho}}$.
Section~\ref{sec_weakall} considers the regime where
$|a_{\sigma}|$ and $|a_{\sigma \sigma'}|$ are much smaller than $a_{\text{ho}}$
and where $1/|k_{\text{so}}|$ is much larger than $a_{\text{ho}}$.
This implies that the energy shifts 
due to the atom-atom interaction and the
spin-orbit coupling are small compared to the harmonic oscillator
energy $E_{\text{ho}}$.
Section~\ref{sec_weaknotall} considers the regime where 
$|a_{\sigma}|$ and $|a_{\sigma \sigma'}|$ 
are not restricted to be small
compared to $a_{\text{ho}}$ and 
where $1/|k_{\text{so}}|$ is much larger than $a_{\text{ho}}$.

\section{Weak atom-atom interaction and
weak spin-orbit coupling}
\label{sec_weakall}
This section pursues a two-step approach: In the first step
(see Sec.~\ref{sec_singleimpurity}), we
determine the 
eigenenergies 
and eigenstates of the single particle
Hamiltonian $H^{(1)}(\vec{r}_j)$ using
Raleigh-Schr\"odinger perturbation theory.
This approach provides a description
for
$|k_{\text{so}}| a_{\text{ho}} \ll 1$.
The perturbative energy and wave function expressions are
given in Eqs.~(\ref{eq_so_pt_series})-(\ref{eq_so_pt4})
and Eqs.~(\ref{eq_wavepert})-(\ref{eq_waveend}), respectively,
and the perturbative energies are compared to
the exact ones in Fig.~\ref{fig_singleparticle}.
In the second step, 
we utilize the eigenstates and eigenenergies 
determined in the first step to treat the interactions
$H_{\text{soc,a}}^{(12)}$ and $H_{\text{soc,soc}}^{(12)}$ 
(see Secs.~\ref{sec_pert_atomimp} and \ref{sec_pert_bb})
perturbatively.
Section~\ref{sec_pert_atomimp} treats the system where one
particle does and the other does not feel the spin-orbit coupling
term. Equations~(\ref{eq_shiftgs})-(\ref{eq_shiftgssplit})
contain the perturbative energy expressions applicable
when the $s$-wave interaction and the spin-orbit coupling term are weak;
these results are validated through comparisons with numerical
results in Figs.~\ref{fig_compare_ai} and \ref{fig_compare_aisplit}.
Section~\ref{sec_pert_bb} considers how the perturbative
energy expressions change when both particles feel the
spin-orbit coupling term.
Equations~(\ref{eq_shifta1}), (\ref{eq_shift1c}) and
(\ref{eq_shift1b}) contain the resulting energy expressions,
and Figs.~\ref{fig_boseshift} and \ref{fig_boseshift2}
respectively illustrate and validate our perturbative results.

\subsection{Single harmonically trapped particle with Rashba coupling}
\label{sec_singleimpurity}
While analytical expressions for the eigenenergies and eigenstates
are reported in the literature
for a single harmonically trapped particle with spin-orbit
coupling of Rashba type~\cite{doucha87, analytic},
we  determine the eigenenergies and eigenfunctions of 
$H^{(1)}(\vec{r}_1)$
perturbatively. Since we are considering a single
particle, we drop the subscript $1$ of the position vector $\vec{r}_1$
in what follows.
We treat the harmonic oscillator Hamiltonian 
$H_{\text{ho}}$ with $\omega_x=\omega_y$
as the unperturbed Hamiltonian and 
$V_{\text{so}}$ as the perturbation.
An analogous approach has been pursued in the quantum
dot literature~\cite{governale02, pi09}. 
An important aspect of our work
is that we go to much higher order in the perturbation series
than earlier work~\cite{governale02}. 
Since $V_{\text{so}}$ is independent of the $z$-coordinate,
it is convenient to employ cylindrical coordinates
$(\rho,\varphi,z)$, where $\rho^2=x^2+y^2$ and
$\tan \varphi=y/x$.
The energy associated with the $z$ coordinate is 
$E_{k_z}=(k_{z}+1/2)\hbar \omega$, where
$k_{z}=0,1,2,\cdots$.

In the following, we focus on the motion
in the $xy$-plane and assume $\omega_x=\omega_y$.
To treat $V_{\text{so}}$ perturbatively,
we write the non-interacting two-dimensional harmonic oscillator 
functions $R_{n_{\rho}, m_l}(\rho) \Phi_{m_l}(\varphi)$
in terms of $\rho$ and $\varphi$,
\begin{eqnarray}
\label{eq_wf_radial}
R_{n_{\rho} m_l}(\rho)=
\sqrt{\frac{2 n_{\rho}!}{a_{\text{ho}}^2 (n_{\rho}+|m_l|)!}}
\left(\frac{\rho}{a_{\text{ho}}} \right)^{|m_l|} \times \nonumber \\
\exp \left(
-\frac{\rho^2}{2 a_{\text{ho}}^2} \right) 
L_{n_{\rho}}^{|m_l|} \left(
\frac{\rho^2}{a_{\text{ho}}^2}
\right),
\end{eqnarray}
where $L_{n_{\rho}}^{|m_l|}$ denotes the associated Laguerre polynomial,
and
\begin{eqnarray}
\label{eq_wf_angle}
\Phi_{m_l}(\varphi)=\frac{1}{\sqrt{2 \pi}} \exp ( \imath  m_l \varphi).
\end{eqnarray}
The principal quantum number $n_{\rho}$ and the projection quantum
number $m_l$ take the values 
$n_{\rho}=0,1,2,\cdots$ and $m_l=0,\pm 1, \pm 2, \cdots$.
The energy associated with the motion in the $x y$-plane is
$E_{n_{\rho},m_l}^{(0)}= (2 n_{\rho} + |m_l|+1) \hbar \omega$.
The unperturbed eigenstates
that account for the pseudo-spin degrees of freedom
can then be written as 
$\psi^{(0)}_{n_{\rho},m_l,m_s}(\rho,\varphi)
=R_{n_{\rho},m_l}(\rho) \Phi_{m_l}(\varphi) |m_s=\pm 1/2 \rangle$.
Since the unperturbed Hamiltonian does not depend on the 
pseudo-spin, each state is two-fold degenerate.
The two-fold degeneracy is not broken by the perturbation $V_{\text{so}}$,
i.e., each exact eigenenergy is two-fold degenerate
due to Kramer's degeneracy 
theorem~\cite{kramer, degeneracytheorem}. This follows from 
the fact that $H^{(1)}$ commutes with the time reversal operator.

When the spin-orbit coupling term is turned on,
the spatial and pseudo-spin degrees of freedom
couple and
$m_l$ and $m_s$ are no longer good quantum numbers.
For non-vanishing $V_{\text{so}}$, $m_j$ with $m_j=m_l+m_s$
is a good quantum number of the Hamiltonian $H^{(1)}$.
The two-fold degeneracy of the unperturbed ground state, e.g.,
arises from the fact that the states with $m_j=1/2$ 
and $m_j=-1/2$ have the same energy.
In general, each unperturbed energy is 
$2E^{(0)}_{n_{\rho},m_l}/(\hbar \omega)$-fold degenerate.
The corresponding wave functions 
are characterized by distinct $m_j$ quantum
numbers.
Since $m_j$ is a good quantum number, the unperturbed wave functions
within a given energy manifold do
not couple.
This implies that we can employ non-degenerate perturbation theory.

\begin{figure}
\includegraphics[angle=0,width=65mm]{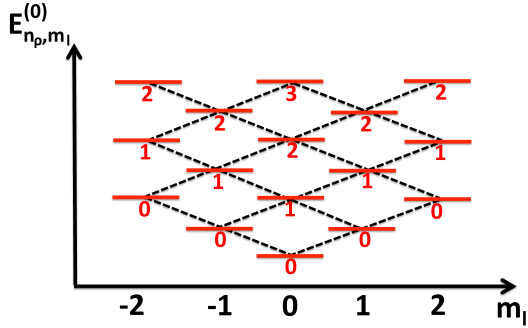}
\caption{(Color online)
Illustration of the selection rules 
[Eqs.~(\ref{eq_matrixelement1}) and (\ref{eq_matrixelement2})]
for a single particle with
spin-orbit coupling of Rashba type.
The horizontal solid lines show the unperturbed
single particle energies $E_{n_{\rho},m_l}^{(0)}$
as a function of the quantum number $m_l$.
The value of $n_{\rho}$ is given below each energy level.
The dotted lines indicate the non-vanishing matrix elements,
i.e., the allowed transitions between unperturbed states
(see text for details).
}
\label{fig_schematic1}
\end{figure}

The perturbation theory expressions (see below)
involve
matrix elements of the type
$\langle \psi^{(0)}_{n_{\rho}',m_l',m_s'} 
| V_{\text{so}} | \psi^{(0)}_{n_{\rho},m_l,m_s} \rangle$.
We find (see also
Refs.~\cite{reimann11, pu12})
\begin{widetext}
\begin{eqnarray}
\label{eq_matrixelement1}
\langle \psi^{(0)}_{n_{\rho},m_l,1/2} 
| V_{\text{so}} | \psi^{(0)}_{n_{\rho}',m_l',-1/2} \rangle=
\frac{\hbar^2 k_{\text{so}}}{m a_{\text{ho}}}
\delta_{m_l,m_l'-1} 
\Bigg\{
\begin{array}{c} 
\left( \delta_{n_{\rho},n_{\rho}'}
\sqrt{n_{\rho}+m_l+1} 
+\delta_{n_{\rho},n_{\rho}'+1} \sqrt{n_{\rho}} \right)
\mbox{ for } m_l \ge 0 \\
\left( -\delta_{n_{\rho},n_{\rho}'}
\sqrt{n_{\rho}+|m_l|} 
- 
\delta_{n_{\rho},n_{\rho}'-1}
\sqrt{n_{\rho}+1} \right)
\mbox{ for } m_l< 0 
\end{array}
\end{eqnarray}
and
\begin{eqnarray}
\label{eq_matrixelement2}
\langle \psi^{(0)}_{n_{\rho},m_l,-1/2} 
| V_{\text{so}} | \psi^{(0)}_{n_{\rho}',m_l',1/2} \rangle=
\frac{\hbar^2 k_{\text{so}}}{m a_{\text{ho}}}
\delta_{m_l,m_l'+1} 
\Bigg\{
\begin{array}{c} 
\left( \delta_{n_{\rho},n_{\rho}'}
\sqrt{n_{\rho}+m_l} 
+\delta_{n_{\rho},n_{\rho}'-1} \sqrt{n_{\rho}+1} \right)
\mbox{ for } m_l > 0 \\
\left( -\delta_{n_{\rho},n_{\rho}'}
\sqrt{n_{\rho}+|m_l|+1} 
- 
\delta_{n_{\rho},n_{\rho}'+1}
\sqrt{n_{\rho}} \right)
\mbox{ for } m_l \le 0 
\end{array}.
\end{eqnarray}
\end{widetext}
The matrix elements for $m_s'=m_s$ vanish. 
This follows from the fact that the spin-orbit coupling term
can be written in terms of the Pauli matrices
$\sigma_x$ and $\sigma_y$, which flip the spin.
The selection rules expressed through the Kronecker delta functions in
Eqs.~(\ref{eq_matrixelement1})
and (\ref{eq_matrixelement2})
are illustrated schematically in Fig.~\ref{fig_schematic1}.
Solid horizontal lines show the unperturbed energies
$E^{(0)}_{n_{\rho},m_l}$ as a function of $m_l$.
The number below each energy level indicates the principal quantum
number $n_{\rho}$. 
Dotted lines indicate 
non-vanishing matrix elements.
It is important to note that the matrix elements are only 
non-zero under certain conditions. For example, let us start
in the $(n_{\rho}',m_l')=(0,0)$ state. 
If $m_s'$ is equal to $1/2$, one can 
reach the $(n_{\rho},m_l)=(0,1)$ state 
(i.e., one can take a step to the right)
but one cannot
reach the $(n_{\rho},m_l)=(0,-1)$ state
(i.e., one cannot take a step to the left).
If $m_s'$ is equal to $-1/2$, in contrast, one can 
reach the $(n_{\rho},m_l)=(0,-1)$ state (i.e.,
one can take
a step to the left)
but 
one cannot reach the $(n_{\rho},m_l)=(0,1)$ state
(i.e., one cannot take a step to the right).

We write the perturbation series as
\begin{eqnarray}
\label{eq_so_pt_series}
E_{n_{\rho},m_l,m_s} 
\approx 
E_{n_{\rho},m_l}^{(0)} + 
\sum_{k=1}^{k_{\text{max}}} \Delta E^{(k)}_{n_{\rho},m_l,m_s},
\end{eqnarray}
where the energy shifts
$\Delta E^{(k)}_{n_{\rho},m_l,m_s}$ are determined by
applying $k$th-order perturbation theory.
Energies $E_{n_{\rho},m_l,m_s}$
with the same $E^{(0)}_{n_{\rho},m_l}$ and $|m_j|$
are degenerate.

\begin{figure}
\vspace*{0.1in}
\includegraphics[angle=0,width=80mm]{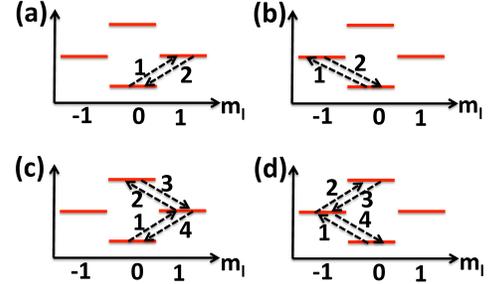}
\caption{(Color online)
Non-zero matrix elements 
for a single particle with
spin-orbit coupling of Rashba type in the
ground state at second- and fourth-order 
perturbation
theory.
Solid horizontal lines show the unperturbed energies $E_{n_{\rho},m_l}^{(0)}$
as a function of $m_l$.
Arrows 
in
panels~(a) and (b) 
show the ``allowed paths'' that contribute
to the energy shift $\Delta E_{0,0,\pm1/2}^{(2)}$.
Arrows in panels~(c) and (d) show the ``allowed paths'' that contribute
to the energy shift $\Delta E_{0,0,\pm1/2}^{(4)}$.
}
\label{fig_schematic2}
\end{figure}

The selection rules discussed above
imply that the first-order energy shift vanishes.
For $k=2$, we have 
\begin{eqnarray}
\label{eq_pt2}
\Delta E_{n_{\rho},m_l,m_s}^{(2)}
=
\sum_{n_{\rho}',m_l',m_s'}
\frac{ |
\langle
\psi_{n_{\rho},m_l,m_s}^{(0)}|V_{\text{so}}|\psi^{(0)}_{n_{\rho}',m_l',m_s'} 
\rangle |^2
} 
{E_{n_{\rho},m_l}^{(0)} - E_{n_{\rho}',m_l'}^{(0)}}
,
\end{eqnarray}
where the sum excludes 
states with 
eigenenergy $E_{n_{\rho},m_l}^{(0)}$.
The matrix elements that
contribute to the second-order perturbation shift
of the ground state are illustrated schematically in 
Figs.~\ref{fig_schematic2}(a) and \ref{fig_schematic2}(b).
The matrix elements give a non-zero contribution
only for $(n_{\rho}',m_l',m_s')=(0,1,-1/2)$
if $m_s=1/2$
and for 
$(n_{\rho}',m_l',m_s')=(0,-1,1/2)$
if $m_s=-1/2$.

We find, in agreement with Refs.~\cite{governale02, zinner12}, 
that the second-order energy shift is given by
\begin{eqnarray}
\label{eq_shift2}
\Delta E^{(2)}_{n_{\rho},m_l,m_s}=
- (1 \pm |m_l|)  E_{\text{so}}
\end{eqnarray}
for $|m_j|=|m_l| \pm 1/2$,
where 
\begin{eqnarray}
E_{\text{so}}= \frac{\hbar^2 k_{\text{so}}^2}{m}.
\end{eqnarray}

We write the $k$th-order perturbation shift
($k$ even) as
\begin{eqnarray}
\label{eq_shiftgeneral}
\Delta E^{(k)}_{n_{\rho},m_l,m_s} =
c_{n_{\rho},m_l,m_s}^{(k)} 
\left( \frac{E_{\text{so}}}{E_{\text{ho}}} \right)^{k/2} E_{\text{ho}}.
\end{eqnarray}
We find that $\Delta E_{n_{\rho},m_l,m_s}^{(k)}=0$ for odd $k$
due to the $m_s$ selection rule.
The $c^{(2)}_{n_{\rho},m_l,m_s}$-coefficients 
can be read off Eq.~(\ref{eq_shift2}). 
Figures ~\ref{fig_schematic2}(c) and \ref{fig_schematic2}(d) illustrate the 
non-zero matrix elements that contribute to the energy shift of the ground
state at fourth-order 
perturbation theory.
Evaluating the perturbation expression, we find
\begin{eqnarray}
\label{eq_so_pt4}
c^{(4)}_{n_{\rho},m_l,m_s}= (2n_{\rho}+|m_l|+1)(1/2 \pm |m_l|)
\end{eqnarray}
for $|m_j|=|m_l| \pm 1/2$.
Table~\ref{table1} summarizes the 
$c_{n_{\rho},m_l,m_s}^{(k)}$ 
coefficients
for $k=2,4,\cdots,12$ for the ground state.

\begin{table}
\caption{Coefficients $c_{0,0,\pm 1/2}^{(k)}$ 
[see Eq.~(\ref{eq_shiftgeneral})]
for a single particle with spin-orbit coupling
of Rashba type.
The coefficients determine the energy shift for the ground state.
}
\label{table1}
\centering
\begin{tabular}{cc|cc}
\hline
\hline
 $k$  &  $c_{0,0,\pm 1/2}^{(k)}$ &
 $k$  &  $c_{0,0,\pm 1/2}^{(k)}$ \\
 \hline
 2 & $-1$ & 8 & $79/72$\\
 4 & $1/2$ & 10 & $-274/135$\\
 6 & $-2/3$ & 12 & $130577/32400$\\
\hline
\hline
\end{tabular}
\end{table}

We developed an analogous scheme to evaluate the 
corrections to the unperturbed wave functions.
We write 
\begin{widetext}
\begin{eqnarray}
\label{eq_wavepert}
\psi_{n_{\rho},m_l,m_s}(\rho,\varphi) 
\approx 
N_{n_{\rho},m_l,m_s}
\Bigg\{
\psi^{(0)}_{n_{\rho},m_l,m_s}(\rho,\varphi) 
+
\sum_{k=1}^{k_{\text{max}}} (k_{\text{so}}a_{\text{ho}})^k
\Bigg[
\sum_{n_{\rho}',m_l',m_s'} d^{(n_{\rho},m_l,m_s,k)}_{n_{\rho}',m_l',m_s'} 
\psi^{(0)}_{n_{\rho}',m_l' ,m_s'}(\rho,\varphi)  \Bigg]
\Bigg\},
\end{eqnarray}
where
the quantum numbers $m_l'$ and $m_s'$ are constrained by
$m_l'+m_s'=m_j$ and where 
the sum excludes states with eigenenergy $E^{(0)}_{n_{\rho},m_l}$.
In Eq.~(\ref{eq_wavepert}),
the normalization constant 
$N_{n_{\rho},m_l,m_s}$ 
can be readily
obtained once the 
$d_{n_{\rho}',m_l',m_s'}^{(n_{\rho},m_l,m_s,k)}$-coefficients
are known,
\begin{eqnarray}
(N_{n_{\rho},m_l,m_s})^{-2}=
1 + 
\sum_{n_{\rho}',m_l',m_s'}
\left[
\sum_{k=1}^{k_{\text{max}}} 
(k_{\text{so}} a_{\text{ho}})^k d_{n_{\rho}',m_l',m_s'}^{(n_{\rho},m_l,m_s,k)}
\right]^2,
\end{eqnarray}
where, as before,
the sum excludes terms corresponding to
eigenenergies $E^{(0)}_{n_{\rho},m_l}$.
For $k=1$ and $2$, we derive general expressions for the
expansion coefficients,
\begin{eqnarray}
d^{(n_{\rho},m_l, \pm1/2,1)}_{n_{\rho}',m_l',m_s'}=
\mp 
\delta_{m_s',\mp1/2}\delta_{m_l',m_l\pm1}
\Bigg\{
\begin{array}{c} 
\left(
\sqrt{n_{\rho}+|m_l|+1}  \delta_{n_{\rho}',n_{\rho}} 
-
\sqrt{n_{\rho}}  
\delta_{n_{\rho}',n_{\rho}-1}
\right)
\mbox{ for } |m_j|=|m_l|+1/2  \\
\left(
\sqrt{n_{\rho}+|m_l|}  \delta_{n_{\rho}',n_{\rho}} 
- 
\sqrt{n_{\rho}+1}  
\delta_{n_{\rho}',n_{\rho}+1} 
\right)
\mbox{ for } |m_j|=|m_l|-1/2
\end{array}
\end{eqnarray}
and
\begin{eqnarray}
\label{eq_waveend}
d^{(n_{\rho},m_l, \pm1/2,2)}_{n_{\rho}',m_l',m_s'}=
\frac{1}{2} 
\delta_{m_s',\pm1/2} \delta_{m_l',m_l}
\bigg(
\sqrt{(n_{\rho}+|m_l|+1)(n_{\rho}+1)}  
\delta_{n_{\rho}',n_{\rho}+1} 
+ 
\sqrt{n_{\rho}(n_{\rho}+|m_l|)}
 \delta_{n_{\rho}',n_{\rho}-1}
 \bigg)
.
\end{eqnarray}

\begin{table}
\caption{Coefficients $d_{n_{\rho}',m_l',m_s'}^{(0,0,\pm 1/2,k)}$ 
[see Eq.~(\ref{eq_wavepert})]
for a single particle with spin-orbit coupling of Rashba type.
The coefficients
determine the wave function corrections for the
ground state with $m_j=\pm 1/2$. 
Columns 2-9 list the coefficients for the 
non-zero $(n_{\rho}',m_l',m_s')$ combinations.
}
\label{table3}
\centering
\begin{tabular}{c|cccccccc}
\hline
\hline
 $k$  &  $(0,\pm 1,\mp 1/2)$ & $(1,0,\pm 1/2)$ & 
$(1, \pm 1,\mp 1/2)$ & $(2,0,\pm 1/2)$ & 
$(2,\pm 1,\mp 1/2)$ & $(3,0,\pm 1/2)$ &
$(3,\pm 1,\mp 1/2)$ & $(4,0,\pm 1/2)$\\
 \hline
 $1$ & $\mp1$ & & & & & & &\\
 $2$ & & $1/2$ & & & & & &\\
 $3$ & $\pm1/2$ & & $\mp\sqrt{2}/6$ & & & & &\\
 $4$ & & $-1/3$ & & $1/12$ & & & &\\
 $5$ & $\mp2/3$ & & $\pm5\sqrt{2}/36$ & & $\mp\sqrt{3}/60$ & & &\\
 $6$ & & $35/72$ & & $-7/90$ & & $1/120$ & &\\
 $7$ & $\pm31/72$ & & $\mp227\sqrt{2}/1080$
        & & $\pm31\sqrt{3}/1800$ & & $\mp1/420$ \\
 $8$ & & $-179/540$ & & $659/5400$ & & $-29/3150$
        & & $1/1680$ \\
\hline
\hline
\end{tabular}
\end{table} 
\end{widetext}
Table~\ref{table3} summarizes the 
$d^{(n_{\rho},m_l,\pm 1/2,k)}_{n_{\rho}',m_l',m_s'}$-coefficients 
for $k=1,2,\cdots,8$
for the ground state, i.e., for $n_{\rho}=0$ and $m_l=0$.

\begin{figure}
\vspace*{+.5cm}
\includegraphics[angle=0,width=70mm]{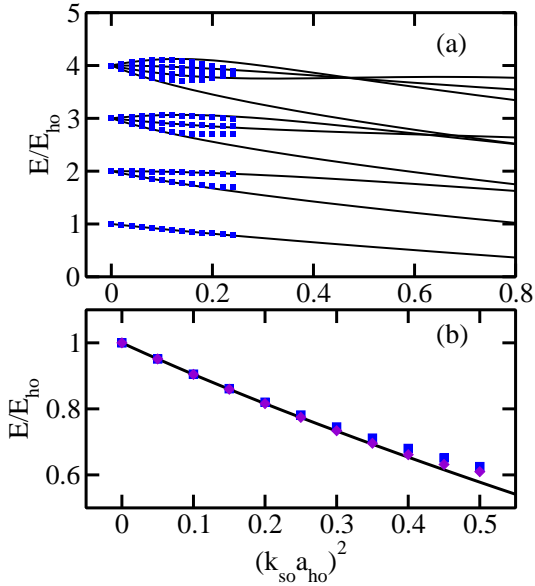}
\vspace*{0.cm}
\caption{(Color online)
Eigenenergies for a single particle with spin-orbit
coupling of Rashba type described by the
Hamiltonian $H^{(1)}$ (the energy associated
with the motion in the $z$ direction has been taken out).
(a) Lines show the numerically determined
(exact) eigenenergies as a function of
$(k_{\text{so}}a_{\text{ho}})^2$. Squares show the energies determined
perturbatively with $k_{\text{max}}=4$.
(b) The ground state energy is shown on an enlarged scale.
Squares and diamonds show the energy determined
perturbatively with $k_{\text{max}}=4$ and $k_{\text{max}}=12$,
respectively.
}
\label{fig_singleparticle}
\end{figure}

To validate our perturbative treatment, we determine the
eigenenergies of $H^{(1)}$ 
($k_{\text{so}} \ge 0$) numerically following the approach
of Ref.~\cite{pu12}.
In the following, we focus on the energies 
associated with the motion in the $x y$-plane and do not
include the energy 
associated with the motion in
the $z$ coordinate.
Solid lines in Fig.~\ref{fig_singleparticle}
show the single particle 
energies 
as a function of $(k_{\text{so}}a_{\text{ho}})^2$. 
For comparison, squares show our perturbative energies $E_{n_{\rho},m_l,m_s}$
with $k_{\text{max}}=4$.
For the excited states shown,
the agreement is excellent for ($k_{\text{so}}a_{\text{ho}})^2 \lesssim 0.1$.
For the ground state [see
also the blow-up in Fig.~\ref{fig_singleparticle}(b)],
the agreement is excellent for $(k_{\text{so}}a_{\text{ho}})^2 \lesssim 0.3$.
Diamonds in Fig.~\ref{fig_singleparticle}(b) show the perturbative energy
for the ground state with $k_{\text{max}}=12$.
It can be seen that the
inclusion of more terms in the perturbation series
improves the agreement with the exact energies
in a narrow $k_{\text{so}} a_{\text{ho}}$ window. As expected, 
as $k_{\text{so}} a_{\text{ho}}$ approaches 1, the perturbative
energy expression fails.

\subsection{Perturbative treatment of $H^{(12)}_{\text{soc,a}}$: one atom with
and one atom without spin-orbit coupling}
\label{sec_pert_atomimp}
 
This section 
accounts for the atom-atom interaction,
modeled using 
$V_{\text{2b}}^{\uparrow}(\vec{r}_{12})=V_{\text{ps}}^{\uparrow}(\vec{r}_{12})$
and
$V_{\text{2b}}^{\downarrow}(\vec{r}_{12})=V_{\text{ps}}^{\downarrow}(\vec{r}_{12})$,
perturbatively.
We first assume $\omega_x=\omega_y=\omega_z$.
We write the unperturbed
two-particle wave function as
a product of the single particle
wave function that accounts for $V_{\text{so}}(\vec{r}_1)$
perturbatively (see Sec.~\ref{sec_singleimpurity})
and the 
single particle harmonic oscillator wave function.
The former describes the motion of the first particle and 
is given by Eq.~(\ref{eq_wavepert}) 
with $\rho=\rho_1$ and $\varphi=\varphi_1$, multiplied
by the one-dimensional harmonic oscillator function 
$g_{k_z}(z_1)$, where $k_z=0,1,\cdots$.
The latter describes the motion of the second particle and
is given by $R_{N_{\rho},M_l}(\rho_2)\Phi_{M_l}(\varphi_2)$
[see Eqs.~(\ref{eq_wf_radial}) and (\ref{eq_wf_angle})],
multiplied
by the one-dimensional harmonic oscillator function 
$g_{K_z}(z_2)$, where $K_z=0,1,\cdots$.
Correspondingly, the unperturbed two-particle energy 
is given by $E_{n_{\rho},m_l,m_s}+(k_z+2 N_{\rho} + |M_l|+K_z+2)\hbar \omega$,
where $E_{n_{\rho},m_l,m_s}$ is given in Eq.~(\ref{eq_so_pt_series}).

Since the atom-atom interaction is spherically symmetric,
unperturbed states with
the same
unperturbed energy 
but different $M_J=m_l+m_s+M_l$ 
do not couple.
To start with, we consider the effect of the 
atom-atom interaction
for case 1a ($a_{\uparrow}=a_{\downarrow}=a_{\text{aa}}$)
on the ground state.
The first-order energy shift 
$\Delta E^{(\text{soc,a},1)}_{\text{gr}, M_J}$ is found by 
``sandwiching''  
$H_{\text{soc,a}}^{(12)}$ between the unperturbed states.
The matrix elements for states with different $m_s$ do not couple.
In the following, we consider the matrix element
that contains $\psi_{0,0,1/2}$ [Eq.~(\ref{eq_wavepert})]; 
considering the matrix element that contains  $\psi_{0,0,-1/2}$ 
yields the same energy shift.
Equation~(\ref{eq_wavepert}) and
Table~\ref{table3}
show that the term proportional to $(k_{\text{so}})^0$
has $m_s=1/2$ while the term proportional to 
$(k_{\text{so}})^1$ has $m_s=-1/2$.
Since these spin states are orthogonal,
the energy shift $\Delta E_{\text{gr}, M_J}^{(\text{soc,a},1)}$
contains a term that is proportional to $a_{\text{aa}}(k_{\text{so}})^0$
(in fact, this is the ``usual'' first-order energy shift one obtains
in the absence of spin-orbit coupling~\cite{busc98}) but
does not contain terms that are proportional to $a_{\text{aa}}k_{\text{so}}$.
Moreover, it can be shown readily that the 
selection rules imply that $\Delta E_{\text{gr}, M_J}^{(\text{soc,a},1)}$
does not contain terms that are proportional to
$a_{\text{aa}}(k_{\text{so}})^k$ with $k$ odd. 

To calculate the coefficient of the term
that is
proportional to $a_{\text{aa}}(k_{\text{so}})^2$,
we have to add up three non-vanishing contributions.
The first contribution comes from the fact that
the normalization constant $N_{0,0,1/2}$
contains a term that is proportional to $(k_{\text{so}})^2$.
The second contribution comes from the fact that
$\psi_{0,0,1/2}$ contains a term that is proportional to
$(k_{\text{so}})^1$, which---when squared---gives a non-vanishing contribution.
The third contribution comes from the fact that
$\psi_{0,0,1/2}$ contains a term that is proportional to
$(k_{\text{so}})^2$, which---when multiplied by the wave function piece 
that is proportional
to $(k_{\text{so}})^0$---gives a non-vanishing contribution.
Evaluating these three finite contributions,
we find that the sum vanishes, i.e.,
the energy shift $\Delta E_{\text{gr}, M_J}^{(\text{soc,a},1)}$
contains no terms that are proportional to $a_{\text{aa}}(k_{\text{so}})^2$.
We refer to the cancellation of this
term as ``accidental''
and note that the coefficient of the $a_{\text{aa}}(k_{\text{so}})^2$
term does, in general,
not vanish when one considers
excited states (see below).

One might ask whether the fact that the perturbative
treatment does not yield a term proportional to
$a_{\text{aa}}(k_{\text{so}})^2$ for the ground state
is a consequence of the azimuthal symmetry. To investigate this question, we 
consider two situations in which the azimuthal symmetry is broken.
We consider the cases where {\em{(i)}} $\omega_x \ne \omega_y$,
and {\em{(ii)}} $\omega_x \ne \omega_y$ and
the Rashba spin-orbit
coupling term is anisotropic, i.e., the 
term proportional to $\partial /\partial x_1$ is multiplied by
a different constant than the term proportional to $\partial /\partial y_1$.
In both cases,
we find that the energy shift of the ground state
does not contain terms that are proportional to $a_{\text{aa}}(k_{\text{so}})^2$.
This shows that the absence of the coupling between the short-range
interaction and the spin-orbit coupling term
for the ground state at order $a_{\text{aa}}(k_{\text{so}})^2$ is not a consequence 
of the azimuthal symmetry. Interestingly, we find
that the term is also absent in the one-dimensional
Hamiltonian with spin-orbit coupling.

Returning to the spherically symmetric harmonic 
confining potential and isotropic Rashba coupling,
we extend the analysis of the ground state
to higher orders in $k_{\text{so}}$. We find
\begin{eqnarray}
\label{eq_shiftgs}
\Delta E^{(\text{soc,a},1)}_{\text{gr},M_J=1/2}= 
\bigg[1+
\frac{1}{4}
(k_{\text{so}}a_{\text{ho}})^4-
\frac{23}{36} 
(k_{\text{so}}a_{\text{ho}})^6+ \nonumber \\
\frac{1513}{1080}
(k_{\text{so}}a_{\text{ho}})^8+
\cdots
\bigg] E_{\text{scatt}},
\end{eqnarray}
where 
\begin{eqnarray}
\label{eq_escale}
E_{\text{scatt}}=\sqrt{\frac{2}{\pi}} \frac{a_{\text{aa}}}{a_{\text{ho}}} E_{\text{ho}}.
\end{eqnarray}
The first term in the square brackets on the right hand side
of Eq.~(\ref{eq_shiftgs}) is the usual $s$-wave 
shift \cite{busc98} and $E_{\text{scatt}}$ can be
interpreted as the ``two-particle" mean-field
shift. The second term gives 
the leading-order coupling between the long-range 
spin-orbit coupling term and the short-range $s$-wave interaction.
Generalizing the above analysis
to excited states with
arbitrary $n_{\rho}$, $m_l$ and $m_s$ but
$N_{\rho}=M_l=K_z=k_z=0$, 
we find that the first-order energy shift is
given by
\begin{eqnarray}
\label{eq_shift_excited}
\Delta E^{(\text{soc,a},1)}_{n_{\rho},m_l,m_s}=
\bigg[
\frac{(2n_{\rho}+|m_l|)!}
{n_{\rho}!(n_{\rho}+|m_l|)!2^{2n_{\rho}+|m_l|}} 
+
\nonumber\\
\bigg(
\frac{(2n_{\rho}+|m_l|+1)!}
{n_{\rho}!(n_{\rho}+|m_l|)!2^{2n_{\rho}+|m_l|}} 
-2n_{\rho}-|m_l|-1
\bigg)\times
\nonumber \\
 (k_{\text{so}}a_{\text{ho}})^2 
 + 
\cdots
\bigg] E_{\text{scatt}}.
\end{eqnarray}

If we 
allow for different scattering
lengths, i.e., if we set  $a_{\uparrow}=a_{\text{aa}}$ 
and $a_{\downarrow}=\eta a_{\text{aa}}$ 
and assume $\eta\neq1$ (case 1b),
then we find that the first-order energy
shift of the unperturbed ground state with $m_s=1/2$ 
($M_J=1/2$)
contains terms proportional to 
$a_{\sigma} (k_{\text{so}})^2$,
\begin{eqnarray}
\label{eq_shiftgssplit}
\Delta E_{\text{gr}, M_J=1/2}^{(\text{soc,a},1)} =
\bigg[1-\frac{1}{2}(1-\eta)(k_{\text{so}} a_{\text{ho}})^2+ \nonumber \\
\frac{1}{12}(13-10\eta)(k_{\text{so}} a_{\text{ho}})^4- \nonumber \\
\frac{1}{180}(441-326\eta)(k_{\text{so}} a_{\text{ho}})^6+ \nonumber \\
\frac{1}{37800}(185677-132722\eta)(k_{\text{so}} a_{\text{ho}})^8+\cdots
\bigg] E_{\text{scatt}}.
\end{eqnarray}
To get the
energy shift $\Delta E_{\text{gr}, M_J=-1/2}^{(\text{soc,a},1)}$
of the unperturbed ground state with $m_s=-1/2$
($M_J=-1/2$),
we replace $\eta$ by $1/\eta$ and $E_{\text{scatt}}$
by $\eta E_{\text{scatt}}$ in Eq.~(\ref{eq_shiftgssplit}).
Equations~(\ref{eq_shiftgs}) and (\ref{eq_shiftgssplit})
show that the interplay between the short-range interaction
and the spin-orbit coupling term is highly tunable.
Specifically, the order at which the coupling arises as well as whether the
interplay leads to a decrease or increase of the energy can be
varied by tuning the $s$-wave scattering lengths.

To validate the perturbative energy
shifts given in Eqs.~(\ref{eq_shiftgs})
and (\ref{eq_shiftgssplit}), we determine the eigenenergies
of the Hamiltonian $H_{\text{soc,a}}$ numerically.
We denote the numerically obtained two-body 
ground state energy by $E_{\text{gr}}^{\text{num}}$.
As discussed in the Appendix, the basis set expansion approach
employs a Gaussian model potential with finite range $r_0$
($r_0=0.02a_{\text{ho}}$);
this implies that a meaningful comparison
of the numerical and perturbative energies has to account for finite-range
effects.
To isolate the interplay between the spin-orbit coupling term
and the $s$-wave interaction, we define the 
energy difference $\Delta E_{\text{gr}}^{\text{num}}$,
\begin{eqnarray}
\label{eq_shift_numerics}
\Delta E_{\text{gr}}^{\text{num}} = 
E_{\text{gr}}^{\text{num}}-E_{\text{gr}}^{s-\text{wave}}-E_{\text{gr}}^{\text{so}}+
3 \hbar \omega.
\end{eqnarray}
Here, $E_{\text{gr}}^{s-\text{wave}}$ denotes the 
two-body ground state energy calculated for
$k_{\text{so}}=0$ using the same finite-range interaction model as used to
calculate $E_{\text{gr}}^{\text{num}}$. The energy $E_{\text{gr}}^{s-\text{wave}}$ is obtained with
high accuracy numerically by solving the one-dimensional
scaled
radial Schr\"odinger equation.
In Eq.~(\ref{eq_shift_numerics}), $E_{\text{gr}}^{\text{so}}$ denotes
the two-body ground state energy calculated in the absence of
the two-body interaction using the same spin-orbit
coupling term as used to calculate
$E_{\text{gr}}^{\text{num}}$. As discussed in the context of Fig.~\ref{fig_singleparticle},
the energy $E_{\text{gr}}^{\text{so}}$ can be obtained with high accuracy
numerically.
For $k_{\text{so}}=0$, our definition implies that
$\Delta E_{\text{gr}}^{\text{num}}$ is equal to zero. For finite $k_{\text{so}}$,
$\Delta E_{\text{gr}}^{\text{num}}$ reflects the interplay between
the spin-orbit coupling term and the $s$-wave interaction.

\begin{figure}
\vspace*{0.3in}
\includegraphics[angle=0,width=65mm]{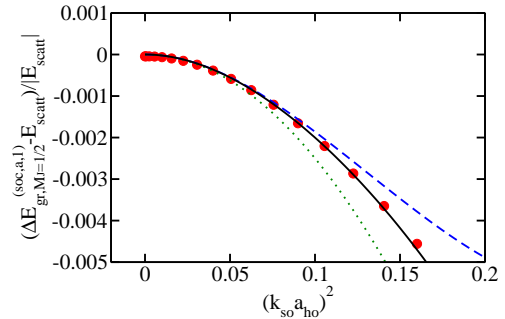}
\caption{(Color online)
Interplay between the 
$s$-wave
interaction and the spin-orbit coupling term
for the ground state for one
atom with and one atom without spin-orbit coupling
(case 1a with 
$a_{\uparrow}=a_{\downarrow}=a_{\text{aa}}=-a_{\text{ho}}/10$).
The lines show the perturbative expression
$(\Delta E_{\text{gr},M_J=1/2}^{(\text{soc,a},1)}-E_{\text{scatt}})/|E_{\text{scatt}}|$,
see Eq.~(\ref{eq_shiftgs}), 
as a function of
$(k_{\text{so}} a_{\text{ho}})^2$.
The dotted, dashed and solid lines show the terms up
to order $(k_{\text{so}}a_{\text{ho}})^4$, $(k_{\text{so}}a_{\text{ho}})^6$ and
$(k_{\text{so}}a_{\text{ho}})^8$, respectively.
For comparison, the circles show
the quantity 
$\Delta E_{\text{gr}}^{\text{num}}/|E_{\text{scatt}}|$,
see Eq.~(\ref{eq_shift_numerics}).
}
\label{fig_compare_ai}
\end{figure}

Figure~\ref{fig_compare_ai}
considers the case where $a_{\uparrow}=a_{\downarrow}=a_{\text{aa}}=-a_{\text{ho}}/10$
(case 1a).
The circles show the quantity 
$\Delta E_{\text{gr}}^{\text{num}}/|E_{\text{scatt}}|$ as a function of $(k_{\text{so}} a_{\text{ho}})^2$.
$E_{\text{gr}}^{\text{num}}$ equals $2.922770(6)\hbar\omega$ for $(k_{\text{so}}a_{\text{ho}})^2=0$
and $2.773036(5)\hbar\omega$ for $(k_{\text{so}} a_{\text{ho}})^2=0.16$
while
$E_{\text{gr}}^{\text{so}}$ equals $2.8506264 \hbar \omega$ for $(k_{\text{so}}a_{\text{ho}})^2=0.16$.
We estimate that the basis set extrapolation 
error for the quantity $\Delta E_{\text{gr}}^{\text{num}}/|E_{\text{scatt}}|$
is less than $7\times10^{-5}$.
For comparison, dotted, dashed and solid lines show the perturbative 
expression 
$(\Delta E_{\text{gr},M_J=1/2}^{(\text{soc,a},1)}-E_{\text{scatt}})/|E_{\text{scatt}}|$,
see Eq.~(\ref{eq_shiftgs}), as a function of $(k_{\text{so}} a_{\text{ho}})^2$
up to order $(k_{\text{so}}a_{\text{ho}})^4$, $(k_{\text{so}}a_{\text{ho}})^6$ and
$(k_{\text{so}}a_{\text{ho}})^8$, respectively. 
The inclusion of more terms in the perturbation series 
systematically improves the agreement with the numerically
determined energy shift.
Equation~(\ref{eq_shiftgs}) accounts for the energy shift proportional to
$a_{\text{aa}}$ but not for energy shifts proportional to $(a_{\text{aa}})^j$
with $j\geq2$.
We find that the leading term in the $(a_{\text{aa}})^2$ series
[see Eq.~(\ref{eq_shiftso4pt}) of Sec.~\ref{sec_swavepert1}]
is, for the $k_{\text{so}}a_{\text{ho}}$ considered in Fig.~\ref{fig_compare_ai},
roughly an order of magnitude
smaller than the smallest contribution included in Eq.~(\ref{eq_shiftgs}).
For example, the energy shift proportional to $(a_{\text{aa}})^2 (k_{\text{so}})^4$
is $-8\times10^{-5}|E_{\text{scatt}}|$ for $(k_{\text{so}} a_{\text{ho}})^2=0.16$.

\begin{figure}
\vspace*{0.3in}
\includegraphics[angle=0,width=65mm]{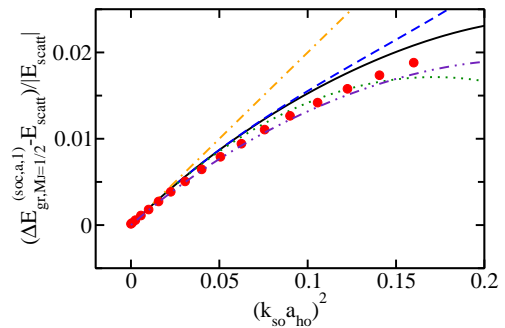}
\caption{(Color online)
Interplay between the 
$s$-wave
interaction and the spin-orbit coupling term
for the ground state for one
atom with and one atom without spin-orbit coupling
(case 1b with $a_{\uparrow}=a_{\text{aa}}=-a_{\text{ho}}/6$
and $a_{\downarrow}=\eta a_{\text{aa}}=-a_{\text{ho}}/10$).
The lines show the perturbative expression
$(\Delta E_{\text{gr},M_J=1/2}^{(\text{soc,a},1)}-E_{\text{scatt}})/|E_{\text{scatt}}|$,
see Eq.~(\ref{eq_shiftgssplit}), 
as a function of
$(k_{\text{so}} a_{\text{ho}})^2$.
The dash-dotted, dotted, dashed and solid
lines show the expression including terms up to
order $(k_{\text{so}} a_{\text{ho}})^2$, $(k_{\text{so}} a_{\text{ho}})^4$, $(k_{\text{so}} a_{\text{ho}})^6$
and $(k_{\text{so}} a_{\text{ho}})^8$, respectively.
The dash-dot-dotted line additionally includes higher-order
corrections in $a_{\text{aa}}$
(see Sec.~\ref{sec_swavepert1} for the derivation of these
higher-order corrections).
For comparison, the circles show the quantity 
$\Delta E_{\text{gr}}^{\text{num}}/|E_{\text{scatt}}|$,
see Eq.~(\ref{eq_shift_numerics}).
}
\label{fig_compare_aisplit}
\end{figure}

Figure~\ref{fig_compare_aisplit}
considers the case where $a_{\uparrow}=a_{\text{aa}}=-a_{\text{ho}}/6$
and $a_{\downarrow}=\eta a_{\text{aa}}=-a_{\text{ho}}/10$ (case 1b).
Circles show the quantity $\Delta E_{\text{gr}}^{\text{num}}$.
As shown in Eq.~(\ref{eq_shiftgssplit}),
the leading-order energy shift that accounts for the interplay
between the spin-orbit coupling term and the $s$-wave interaction
is proportional to $a_{\text{aa}}(k_{\text{so}})^2$
(see the dash-dotted line in Fig.~\ref{fig_compare_aisplit}).
When terms up to order $(k_{\text{so}}a_{\text{ho}})^8$ are included
(see the solid line in Fig.~\ref{fig_compare_aisplit}),
the first-order perturbation theory shift
proportional to $a_{\text{aa}}$
agrees reasonably well with the numerical data.
Since $|a_{\text{aa}}|/a_{\text{ho}}$ is appreciable
($a_{\text{aa}}/a_{\text{ho}}=-1/6$), higher-order corrections in
$a_{\text{aa}}$ are non-negligible. The dash-dot-dotted line
in Fig.~\ref{fig_compare_aisplit},
which additionally includes higher-order corrections in $a_{\text{aa}}$
[see Eq.~(\ref{eq_shiftso4}) in Sec.~\ref{sec_swavepert1}],
notably improves the agreement with the numerically 
determined energy shift.

Figures \ref{fig_compare_ai} and \ref{fig_compare_aisplit}
report the energy shift that reflects the interplay
between the spin-orbit coupling term
and the $s$-wave interaction
in terms of the quantity $|E_{\text{scatt}}|$, i.e., in terms of the
absolute value of the leading-order mean-field shift.
In Figs.~\ref{fig_compare_ai} and \ref{fig_compare_aisplit},
the quantity 
$|(\Delta E_{\text{gr},M_J=1/2}^{(\text{soc,a},1)}-E_{\text{scatt}})/E_{\text{scatt}}|$
is smaller than $5\times 10^{-3}$ and $4\times 10^{-2}$,
respectively, implying that the energy shift
due to the interplay between the spin-orbit coupling term and 
the $s$-wave interaction is respectively less than a
percent and a few percent of the mean-field shift.
While these effects are small, they can potentially
be measured in ``quantum phase revival experiments" analogous to
those for few-atom systems in an optical lattice~\cite{bloch10}.
In that work, it was possible to deduce the effective
three-body interaction energy, which was
measured to be roughly 10 times
smaller in absolute value than the effective two-body interaction energy.
Moreover, the effective four-body energy was measured to be
roughly a factor of 100 smaller than the 
effective two-body interaction. 
To probe the interplay between the spin-orbit coupling term
and the $s$-wave interaction experimentally, one would
compare the oscillation periods in revival experiments with and
without spin-orbit coupling.

The treatment discussed in this section
can, in principle, be extended to
second- and higher-order perturbation theory.
However, the use of the interaction model
$V_{\text{ps}}^{\sigma}(\vec{r}_{12})$ 
gives rise, at second- and higher-order perturbation theory,
to divergencies that need to be removed through application of a
renormalization scheme. Although this can be done via
standard techniques (see, e.g., Refs.~\cite{NJP1,NJP2}),
we find it easier to determine the energy shifts that
are proportional to
$(a_{\text{aa}})^2 (k_{\text{so}})^2$  and $(a_{\text{aa}})^2 (k_{\text{so}})^4$ 
by an approach that builds on the exact two-particle
$s$-wave solution
(see Sec.~\ref{sec_weaknotall}).

The key points of this section are:
\begin{itemize}
\item
For the ground state manifold, the perturbative energy shifts 
contain even but not odd powers of $k_{\text{so}}a_{\text{ho}}$.
\item
For $a_{\uparrow}=a_{\downarrow}=a_{\text{aa}}$ ($\eta=1$), 
the energy shift proportional to
$a_{\text{aa}}(k_{\text{so}})^2$ vanishes for the ground state.
This finding does not only hold for isotropic Rashba coupling and
isotropic traps, but also for anisotropic Rashba coupling
and/or anisotropic harmonic traps. In general,
the energy shift proportional to
$a_{\text{aa}}(k_{\text{so}})^2$ does not vanish for excited states
[see Eqs.~(\ref{eq_shiftgs}) and (\ref{eq_shift_excited})].
\item
For $a_{\uparrow}=a_{\text{aa}}\neq a_{\downarrow}$ ($\eta\neq1$), 
the leading-order energy shifts of the states
in the lowest energy manifold due to the interplay between the
spin-orbit coupling term and the $s$-wave interaction are
proportional to $a_{\text{aa}}(k_{\text{so}})^2$.
\end{itemize}

\subsection{Perturbative treatment of $H^{(12)}_{\text{soc,soc}}$: Two
particles with spin-orbit coupling}
\label{sec_pert_bb}
This section considers the situation where both particles
feel the Rashba spin-orbit coupling.
Throughout, we assume $\omega_x=\omega_y=\omega_z$.
We write the unperturbed two-particle wave function
as a product of two single-particle wave functions, which account
for the spin-orbit coupling terms 
$V_{\text{so}}(\vec{r}_1)$ and $V_{\text{so}}(\vec{r}_2)$ perturbatively.
For concreteness, we focus on the ground state manifold
that consists of the
unperturbed wavefunctions $\Psi^{(0)}_{m_{s1},m_{s2}}$,
where
$\Psi^{(0)}_{m_{s1},m_{s2}}=\psi_{0,0,m_{s1}}(\rho_1,\varphi_1)g_{0}(z_1) \psi_{0,0,m_{s2}}(\rho_2,\varphi_2)g_{0}(z_2)$ and
$(m_{s1},m_{s2})=(1/2,1/2),
(-1/2,-1/2),
(1/2,-1/2)$
and
$(-1/2,1/2)$.
As before, $\psi$ is given by Eq.~(\ref{eq_wavepert}) and $g_0$ denotes the
one-dimensional harmonic oscillator function with energy $\hbar \omega/2$.
The four degenerate unperturbed wave functions are
eigenstates of the total $J_z$ operator with
eigenvalue $\hbar M_J$
($M_J=m_{l1}+m_{s1}+m_{l2}+m_{s2}$
or, equivalently, $M_J=m_{j1}+m_{j2}$), 
where $M_J=1,-1,0$ and $0$, respectively.
Since $M_J$ is a good quantum number, the perturbation $H^{(12)}_{\text{soc,soc}}$
only couples states with the same $M_J$.
In what follows, we use 
$V_{\text{2b}}^{\sigma \sigma'}(\vec{r}_{12})= V_{\text{ps}}^{\sigma \sigma'}(\vec{r}_{12})$
and treat
$H_{\text{soc,soc}}^{(12)}$
in first-order perturbation theory.

We start by considering case 2d ($\zeta, \eta \neq 1$ and 
$\zeta \neq \eta$).
For the state with $M_J=1$, the 
first-order energy 
shift in the scattering length
is given by
\begin{eqnarray}
\label{eq_shifta1}
\Delta E^{(\text{soc,soc},1)}_{\text{gr},M_J=1(S)}=
\bigg[
1
-\left(1
- \eta
\right)
(k_{\text{so}}a_{\text{ho}})^2 
+
\nonumber \\
\frac{1}{6}\left(16+3\zeta
-16 \eta
 \right)
(k_{\text{so}}a_{\text{ho}})^4 -
\nonumber \\
\frac{1}{90}\left(606+165\zeta-656\eta\right)
(k_{\text{so}}a_{\text{ho}})^6 +\nonumber \\
\frac{1}{37800}
\left(589229+215250\zeta-693844\eta\right)
(k_{\text{so}}a_{\text{ho}})^8+
\nonumber \\
\cdots
\bigg]
E_{\text{scatt}},
\end{eqnarray}
where $E_{\text{scatt}}$ is defined in Eq.~(\ref{eq_escale}).
The subscript ``$(S)$"
indicates that the corresponding eigenstate
is symmetric under the exchange of particles 1 and 2.
Similarly, for the state with $M_J=-1$,
the first-order energy shift 
$\Delta E^{(\text{soc,soc},1)}_{\text{gr},M_J=-1(S)}$
is given by
Eq.~(\ref{eq_shifta1}) with
$\zeta$ replaced by $1/\zeta$,
$\eta$ replaced by $\eta/\zeta$
and $E_{\text{scatt}}$ replaced by $\zeta E_{\text{scatt}}$.

We find that the two states with $M_J=0$ couple. This means that we 
have to
employ first-order degenerate perturbation theory.
The diagonal elements 
$\langle
\Psi^{(0)}_{1/2,-1/2}|H^{(12)}_{\text{soc,soc}}|\Psi^{(0)}_{1/2,-1/2}
\rangle$
and
$\langle 
\Psi^{(0)}_{-1/2,1/2}|H^{(12)}_{\text{soc,soc}}|\Psi^{(0)}_{-1/2,1/2}
\rangle$
of the perturbation matrix are given by 
Eq.~(\ref{eq_shifta1}) with
$\zeta$ replaced by $1$,
$\eta$ replaced by $(1/\eta+\zeta/\eta)/2$
and $E_{\text{scatt}}$ replaced by $\eta E_{\text{scatt}}$.
For the off-diagonal elements, we find
\begin{eqnarray}
\langle \Psi^{(0)}_{1/2,-1/2} |
H^{(12)}_{\text{soc,soc}} | \Psi_{-1/2,1/2}^{(0)} \rangle=
\nonumber \\
\langle \Psi^{(0)}_{-1/2,1/2} |
H^{(12)}_{\text{soc,soc}} | \Psi_{1/2,-1/2}^{(0)} \rangle=
 \nonumber \\
\bigg[
\frac{1}{2}
(k_{\text{so}}a_{\text{ho}})^2 - 
\frac{4}{3}
(k_{\text{so}}a_{\text{ho}})^4 +
\frac{164}{45}
(k_{\text{so}}a_{\text{ho}})^6  -\nonumber \\
\frac{173461}{18900}
(k_{\text{so}}a_{\text{ho}})^8+
\cdots \bigg]\left(1+\zeta-2\eta\right) 
E_{\text{scatt}}.
\end{eqnarray}
Diagonalizing the 
$2\times2$ perturbation matrix, we find 
\begin{eqnarray}
\label{eq_shift1c}
\Delta E^{(\text{soc,soc},1)}_{\text{gr},M_J=0(S)}=
\bigg[
\eta
+ \left(1+\zeta
- 2\eta
\right)
(k_{\text{so}}a_{\text{ho}})^2 -
\nonumber \\
\frac{1}{6}
\left(16+16\zeta
-35 \eta
\right)
(k_{\text{so}}a_{\text{ho}})^4 +
\nonumber \\
\frac{1}{90}
\left(656+656\zeta
-1427 \eta
\right)
(k_{\text{so}}a_{\text{ho}})^6 -\nonumber \\
\frac{1}{37800}
\left(693844+693844\zeta-1498323 \eta\right)
(k_{\text{so}}a_{\text{ho}})^8
\nonumber \\
+\cdots
\bigg] 
E_{\text{scatt}}
\end{eqnarray}
and
\begin{eqnarray}
\label{eq_shift1b}
\Delta E^{(\text{soc,soc},1)}_{\text{gr},M_J=0(A)}=
\bigg[
1+\frac{1}{2} 
(k_{\text{so}}a_{\text{ho}})^4
-\frac{23}{18}
(k_{\text{so}}a_{\text{ho}})^6 +\nonumber \\
\frac{3161}{1080}
(k_{\text{so}}a_{\text{ho}})^8
+\cdots
\bigg] \eta
E_{\text{scatt}}.
\end{eqnarray}
The corresponding eigenstates are
$(\Psi^{(0)}_{1/2,-1/2}+\Psi^{(0)}_{-1/2,1/2})/\sqrt{2}$
and
$(\Psi^{(0)}_{1/2,-1/2}-\Psi^{(0)}_{-1/2,1/2})/\sqrt{2}$,
respectively. The
former state is symmetric under the exchange of
particles 1 and 2, while the latter is
anti-symmetric under the exchange of particles 1 and 2.
The symmetry of the states is indicated by the
subscripts ``$(S)$'' and ``$(A)$'' in
Eqs.~(\ref{eq_shift1c}) and  (\ref{eq_shift1b}), respectively.

Our calculations imply that the ground state manifold for two identical bosons
contains three states, whose energy shifts 
are given by Eq.~(\ref{eq_shifta1}),
Eq.~(\ref{eq_shifta1}) with the substitutions discussed below
the equation and Eq.~(\ref{eq_shift1c}).
For two identical fermions, the ground state manifold contains a
single state, whose
energy shift is given by Eq.~(\ref{eq_shift1b}).
As expected, the energy shift corresponding to the anti-symmetric
state is independent of $a_{\uparrow \uparrow}$
and $a_{\downarrow \downarrow}$.
Although our interaction model allows for $s$-wave scattering in all
four channels (up-up, down-down, up-down, down-up),
the anti-symmetry of the wave function ``turns off" the
interactions in the up-up and down-down channels, yielding
an energy shift that is fully determined by
$a_{\uparrow \downarrow}=a_{\downarrow \uparrow}=\eta a_{\text{aa}}$.
The energy shifts corresponding  to the three symmetric states contain a
term proportional to $(k_{\text{so}}a_{\text{ho}})^2$ while
the energy shift corresponding to the anti-symmetric state does not contain a term proportional to $(k_{\text{so}}a_{\text{ho}})^2$.

While our derivation above assumed $\zeta, \eta \neq1$ and $\zeta \neq \eta$
(case 2d), the energy shifts for cases 2a-2c can be obtained by
taking the appropriate limits in Eqs.~(\ref{eq_shifta1})-(\ref{eq_shift1b}).
In the limit that $\zeta=1$ and $\eta\neq1$ (case 2b),
the energy shifts of the two $|M_J|=1$ states with bosonic exchange
symmetry are equal
to each other and contain terms proportional 
to $a_{\text{aa}}(k_{\text{so}})^2$. The $M_J=0$ 
state with bosonic exchange symmetry also contains 
a shift proportional to $a_{\text{aa}}(k_{\text{so}})^2$.
In the limit that $\zeta\neq1$ and $\eta=1$ (case 2c),
the energy shift of the $M_J=1$ state contains no term
proportional to $a_{\text{aa}}(k_{\text{so}})^2$ while
the energy shift of the 
$M_J=-1$ and $M_J=0$ states with
bosonic exchange symmetry contain terms proportional to
$a_{\text{aa}}(k_{\text{so}})^2$. 
In the limit that $\zeta=\eta=1$ (case 2a),
the degeneracy of the unperturbed states is preserved, i.e.,
the four energy shifts of the ground state manifold 
are all equal to each other
and given by Eq.~(\ref{eq_shift1b}). In this case,
the energy shift of the 
ground state
contains no terms that are 
proportional to $a_{\text{aa}}(k_{\text{so}})^2$.
Interestingly, the energy shift given in Eq.~(\ref{eq_shift1b})
is nearly identical to the shift given in Eq.~(\ref{eq_shiftgs})
for the two-atom system where only one of the particles feels
the spin-orbit coupling. Specifically, terms proportional
to $(k_{\text{so}}a_{\text{ho}})^4$ and $(k_{\text{so}}a_{\text{ho}})^6$
differ by a factor of 2, reflecting the fact that the interplay
between the spin-orbit coupling term and the $s$-wave interaction scales
with the number of particles that feel the spin-orbit coupling
term. At order $(k_{\text{so}}a_{\text{ho}})^8$, the two expressions differ by a 
factor different from 2, indicating that the interplay
between the spin-orbit coupling term and the $s$-wave
interaction is not simply additive at higher orders.

\begin{figure}
\vspace*{0.3in}
\includegraphics[angle=0,width=65mm]{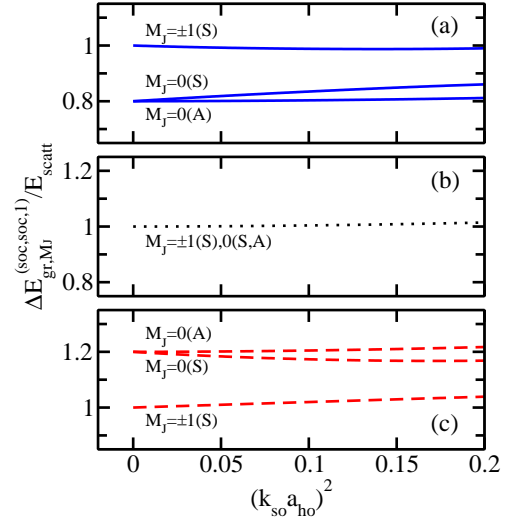}
\caption{(Color online)
First-order energy shift 
$\Delta E^{(\text{soc,soc},1)}_{\text{gr},M_J}$
for the ground state manifold for two 
identical particles with spin-orbit coupling
(case 2a with $\zeta=1$ and $\eta=1$, and
case 2b with $\zeta=1$ and $\eta\neq1$).
Solid, dotted and dashed lines show the quantity
$\Delta E^{(\text{soc,soc},1)}_{\text{gr},M_J}/E_{\text{scatt}}$
[see Eqs.~(\ref{eq_shifta1}), (\ref{eq_shift1c})
and (\ref{eq_shift1b})]
as a function of $(k_{\text{so}} a_{\text{ho}})^2$ 
for (a) $\eta=0.8$, (b) $\eta=1$ and (c) $\eta=1.2$, respectively.
The energy levels are labeled by the $M_J$ quantum
number and the exchange symmetry ($S/A$) of the
corresponding states. 
}
\label{fig_boseshift}
\end{figure}

To illustrate the behavior of the energy level structure of the 
ground state manifold for two identical particles,
we focus on systems with $\zeta=1$.
Lines in Fig.~\ref{fig_boseshift}
show the quantity $\Delta E_{\text{gr},M_J}^{(\text{soc,soc},1)}/E_{\text{scatt}}$
as a function of $(k_{\text{so}} a_{\text{ho}})^2$
for (a) $\eta = 0.8$, (b) $\eta=1$ and (c) 
$\eta=1.2$.
For $\eta=1$ [case 2a, Fig.~\ref{fig_boseshift}(b)], 
the four energy shifts
for the states with $M_J=0$ and $\pm1$ are the same 
(see discussion above).
For $\eta=0.8$ [case 2b, Fig.~\ref{fig_boseshift}(a)],
the $M_J=0$ state with fermionic exchange symmetry 
has lower energy if $a_{\text{aa}}>0$ while
the two-fold degenerate $|M_J|=1$ states with bosonic
exchange symmetry
have lower energy if $a_{\text{aa}}<0$.
For  $\eta=1.2$ [Fig.~\ref{fig_boseshift}(c)],
the two-fold degenerate $|M_J|=1$ states 
with bosonic exchange symmetry
have lower energy if $a_{\text{aa}}>0$ while
the $M_J=0$ state with fermionic exchange symmetry
has lower energy if $a_{\text{aa}}<0$.

\begin{figure}
\vspace*{0.3in}
\includegraphics[angle=0,width=65mm]{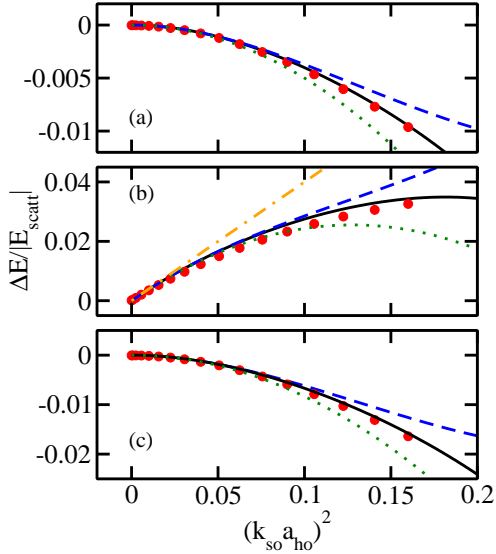}
\caption{(Color online)
Interplay between the 
$s$-wave
interaction and the spin-orbit coupling term
for the ground state manifold for two
atoms with spin-orbit coupling.
(a) The dotted, dashed and solid lines show the expression
$(\Delta E_{\text{gr},M_J=1(S)}^{(\text{soc,soc},1)}-E_{\text{scatt}})/|E_{\text{scatt}}|$,
 Eq.~(\ref{eq_shifta1}),
for the lowest energy state including terms up to order
$(k_{\text{so}}a_{\text{ho}})^4$, $(k_{\text{so}}a_{\text{ho}})^6$ and $(k_{\text{so}}a_{\text{ho}})^8$, 
respectively,
as a function of
$(k_{\text{so}} a_{\text{ho}})^2$ for
case 2a with $a_{\text{aa}}=-a_{\text{ho}}/10$, $\zeta=1$ and $\eta =1$.
(b) The dash-dotted, dotted, dashed and solid lines 
show the expression
$(\Delta E_{\text{gr},M_J=1(S)}^{(\text{soc,soc},1)}-E_{\text{scatt}})/|E_{\text{scatt}}|$,
Eq.~(\ref{eq_shifta1}),
for the lowest energy state
including terms up to order
$(k_{\text{so}}a_{\text{ho}})^2$, $(k_{\text{so}}a_{\text{ho}})^4$, 
$(k_{\text{so}}a_{\text{ho}})^6$ and $(k_{\text{so}}a_{\text{ho}})^8$, 
respectively,
as a function of
$(k_{\text{so}} a_{\text{ho}})^2$ for case 2b with
$a_{\text{aa}}=-a_{\text{ho}}/6$, $\zeta=1$ and $\eta a_{\text{aa}}=-a_{\text{ho}}/10$.
(c) The dotted, dashed and solid lines show the expression
$(\Delta E_{\text{gr},M_J=0(A)}^{(\text{soc,soc},1)}-\eta E_{\text{scatt}})/|E_{\text{scatt}}|$,
Eq.~(\ref{eq_shift1b}),
for the ground state including terms up to order
$(k_{\text{so}}a_{\text{ho}})^4$, $(k_{\text{so}}a_{\text{ho}})^6$ and $(k_{\text{so}}a_{\text{ho}})^8$, 
respectively,
as a function of
$(k_{\text{so}} a_{\text{ho}})^2$ for 
case 2b with $a_{\text{aa}}=-a_{\text{ho}}/10$, $\zeta=1$ 
and $\eta a_{\text{aa}}=-a_{\text{ho}}/6$.
For comparison, the circles show the quantity
$\Delta E_{\text{gr}}^{\text{num}}/|E_{\text{scatt}}|$,
see Eq.~(\ref{eq_shift_numerics}).
}
\label{fig_boseshift2}
\end{figure}

Figure~\ref{fig_boseshift2} compares the perturbative predictions (lines)
with our numerical basis set expansion results (circles).
Figure~\ref{fig_boseshift2}(a) shows an example for 
$a_{\text{aa}}=-a_{\text{ho}}/10$ and $\zeta=\eta=1$ (case 2a).
In this case, the ground state is four-fold degenerate
and the term proportional to $a_{\text{aa}}(k_{\text{so}})^2$
is absent. Figure~\ref{fig_boseshift2}(b) shows the case where
$a_{\text{aa}}=-a_{\text{ho}}/6$, $\zeta=1$ and $\eta a_{\text{aa}}=-a_{\text{ho}}/10$ (case 2b).
According to the analysis
above, the lowest energy state is two-fold degenerate ($|M_J|=1$)
and possesses bosonic exchange symmetry. 
The leading-order energy shift is proportional
to  $a_{\text{aa}}(k_{\text{so}})^2$. 
Figure~\ref{fig_boseshift2}(c) shows the case where
$a_{\text{aa}}=-a_{\text{ho}}/10$, $\zeta=1$ and $\eta a_{\text{aa}}=-a_{\text{ho}}/6$ (case 2b). 
According to the analysis above,
the lowest energy state is one-fold degenerate
($M_J=0$) and possesses fermionic exchange symmetry.
The energy shift is given by Eq.~(\ref{eq_shift1b}), where the term
proportional to $a_{\text{aa}}(k_{\text{so}})^2$
is again absent \cite{footnote}. 
Figure~\ref{fig_boseshift2} demonstrates 
excellent agreement between the perturbative
predictions and our numerical results for all cases.

The key points of this section are:
\begin{itemize}
\item
For the ground state manifold, the perturbative energy shifts 
contain even but not odd powers of $k_{\text{so}}a_{\text{ho}}$.
\item
For two identical bosons, the energy shift proportional
to $a_{\text{aa}}(k_{\text{so}})^2$ is non-zero for the ground state unless
the scattering lengths in the four spin channels
are such that $a_{\uparrow\uparrow}=a_{\uparrow\downarrow}
=a_{\downarrow\uparrow}$ 
($1-\eta=0$) or 
$a_{\uparrow\uparrow}+a_{\downarrow\downarrow}
-2a_{\uparrow\downarrow}=0$
($1+\zeta-2\eta=0$).
\item
For two identical fermions, the energy shift of the ground
state does not contain a term proportional to $a_{\text{aa}}(k_{\text{so}})^2$.
\end{itemize}

\section{Arbitrary atom-atom scattering length and
weak spin-orbit coupling}
\label{sec_weaknotall}
This section takes advantage of the fact that the 
solution for two particles without
spin-orbit coupling under external spherically symmetric
confinement interacting through the regularized pseudopotential
$V_{\text{ps,reg}}(\vec{r}_{12})$ is
known in compact analytical form for arbitrary
$s$-wave scattering length~\cite{busc98}.
Motivated by this, we
treat the spin-orbit coupling perturbatively.
Section~\ref{sec_swavetwobody} reviews the solution
for two particles without spin-orbit coupling.
The two-particle energy spectrum for $k_{\text{so}}=0$
is shown in Fig.~\ref{fig_busch}(b) as 
a function of the inverse of the $s$-wave scattering length.
Sections~\ref{sec_swavepert1}-\ref{sec_swavepert2}
discuss, using the exact
two-body $s$-wave solution, the perturbative treatment
of 
$V_{\text{so}}(\vec{r}_1)$
and $V_{\text{so}}(\vec{r}_1)+V_{\text{so}}(\vec{r}_2)$.
Section \ref{sec_swavepert1} treats the system
where one particle does and the other does not feel the
spin-orbit coupling term assuming small $|k_{\text{so}}|a_{\text{ho}}$
but arbitrary $s$-wave scattering lengths.
Equations (\ref{eq_shift_soca1})-(\ref{eq_d2pt})
contain the resulting perturbative energy expressions,
which are applicable when the states in the manifold studied
are not degenerate with other states.
Figures \ref{fig_coupling}/\ref{fig_d2}
and \ref{fig_compare_aiunit}/\ref{fig_compare_aiunitni}
respectively illustrate and validate these perturbative 
results. The regime where states in the manifold
studied are degenerate with other states is
studied in Sec.~\ref{sec_degenerate} via
near-degenerate perturbation theory for
selected examples (see Fig.~\ref{fig_deg} for an illustration
of the results). Lastly, Sec.~\ref{sec_swavepert2}
treats the system where both particles feel the 
spin-orbit coupling term assuming small $|k_{\text{so}}|a_{\text{ho}}$
but arbitrary $s$-wave scattering lengths.
Equations (\ref{eq_shift_socsoc1})-(\ref{eq_shift_socsoc3})
contain the resulting perturbative energy expressions and 
Fig.~\ref{fig_compare_socsoc} validates these results through
comparison with ``exact" numerical energies. 

\subsection{Two-body wave function for
arbitrary atom-atom scattering length}
\label{sec_swavetwobody}
Throughout, we assume $\omega_x=\omega_y=\omega_z$.
In this case, the 
two-body solution for 
two particles without spin-orbit coupling and arbitrary
$a_{\text{aa}}$ is most conveniently
written in terms of the relative distance vector
$\vec{r}_{12}$ and the center of mass vector
$\vec{R}_{12}$, $\vec{R}_{12}=(\vec{r}_1+\vec{r}_2)/2$.
Specifically, the total two-body
wave function 
can be written as a product
of the relative wave function $\psi_{q_{\text{rel}},l_{\text{rel}},m_{\text{rel}}}^{\text{rel}}$
and the center of mass 
wave function
$\psi^{\text{cm}}_{N_{\text{cm}},M_{\text{cm}}, K_{\text{cm}}}$, and the two-particle energy is given by the
sum of the relative and center of mass contributions.

\begin{figure}
\vspace*{0.3in}
\includegraphics[angle=0,width=65mm]{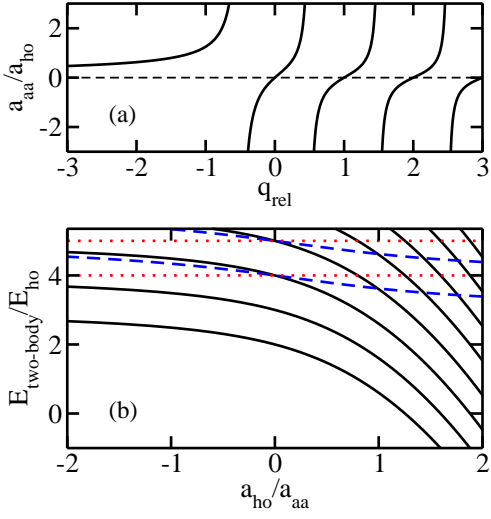}
\caption{(Color online)
Energy spectrum for two particles without
spin-orbit coupling and arbitrary $a_{\text{aa}}$.
(a) The solid lines show $a_{\text{aa}}/a_{\text{ho}}$ as a function of
the non-integer quantum number $q_{\text{rel}}$.
$q_{\text{rel}}$ takes on integer values when $a_{\text{aa}}=0$
and half-integer values when $1/a_{\text{aa}}=0$.
The dashed line shows the ``zero line".
(b) Lines show the two-body energy
$E_{\text{two-body}}$ as a function of
$a_{\text{ho}}/a_{\text{aa}}$. The solid and dashed lines show the
energies of states with
$l_{\text{rel}}=0$ while the dotted lines show the
energies of states with
$l_{\text{rel}}>0$.
The lowest solid and lowest dashed lines
show energies of states without center of mass
excitations.
}
\label{fig_busch}
\end{figure}

The relative wave function is obtained by
solving the relative Schr\"odinger equation
using spherical coordinates. For relative orbital angular momentum $l_{\text{rel}}=0$
and corresponding projection quantum number $m_{\text{rel}}=0$, 
the relative wave function
reads~\cite{busc98}
\begin{eqnarray}
\psi_{q_{\text{rel}}, 0,0}^{\text{rel}} (\vec{r}_{12})=
\frac{N_{q_{\text{rel}}}}{\sqrt{4 \pi}} 
U \left(
-q_{\text{rel}},\frac{3}{2},\frac{1}{2}\left[ \frac{r_{12}}{a_{\text{ho}}} \right]^2
\right) \times
\nonumber \\
 e^{-\frac{1}{4} \left( \frac{r_{12}}{a_{\text{ho}}} \right)^2},
\end{eqnarray}
where $U$ is the confluent hypergeometric function
and $N_{q_{\text{rel}}}$ is
the normalization constant [see Eq.~(B3) of Ref.~\cite{daily12}
for an explicit expression for $N_{q_{\text{rel}}}$; see also Ref.~\cite{busc98}].
The allowed non-integer quantum numbers $q_{\text{rel}}$ 
are obtained by solving the transcendental
equation~\cite{busc98}
\begin{eqnarray}
\label{eq_transcendental}
\frac{\sqrt2 \Gamma(-q_{\text{rel}})}{\Gamma(-q_{\text{rel}}-1/2)} = \frac{a_{\text{ho}}}{a_{\text{aa}}}.
\end{eqnarray}
The relative $l_{\text{rel}}=0$ eigenenergies are given by
$(2 q_{\text{rel}} + 3/2) \hbar \omega$.
Figure \ref{fig_busch}(a) illustrates the relationship between
$q_{\text{rel}}$ and $a_{\text{aa}}$.
In the non-interacting regime, e.g., one finds
$q_{\text{rel}}=0,1,2,\cdots$;
for $|a_{\text{aa}}|=\infty$, in contrast, one finds $q_{\text{rel}}=-1/2,1/2,3/2,\cdots$.
The relative 
states with $l_{\text{rel}}>0$ are not affected by the $s$-wave interaction and
are given by the three-dimensional harmonic oscillator 
states with quantum numbers $n_{\text{rel}}$, $l_{\text{rel}}$ and $m_{\text{rel}}$.

The center of mass wave functions 
$\psi^{\text{cm}}_{N_{\text{cm}},M_{\text{cm}}, K_{\text{cm}}}$
coincide with the three-dimensional 
harmonic oscillator states.
Since
the center of mass wave functions are conveniently written
in cylindrical coordinates, we use the 
quantum numbers $N_{\text{cm}}$, $M_{\text{cm}}$ and $K_{\text{cm}}$ with
$N_{\text{cm}}$ and $K_{\text{cm}}=0,1,\cdots$ and $M_{\text{cm}}=0,\pm1,\cdots$ as labels.
Figure \ref{fig_busch}(b) shows the two-particle energy spectrum
as a function of $a_{\text{ho}}/a_{\text{aa}}$.
Energy levels corresponding to states with $l_{\text{rel}}=0$ are
shown by solid and dashed lines while those
corresponding to $l_{\text{rel}}>0$ are shown by dotted lines.
The following sections investigate how the spin-orbit coupling
term modifies the energy spectrum shown in Fig.~\ref{fig_busch}(b).

\subsection{Perturbative treatment of $V_{\text{so}}(\vec{r}_1)$:
One atom with and one atom without
spin-orbit coupling}
\label{sec_swavepert1}
To treat the spin-orbit term $V_{\text{so}}(\vec{r}_1)$
perturbatively,
we transform it to relative and center of mass coordinates,
\begin{eqnarray}
V_{\text{so}}(\vec{r}_1)=
V_{\text{so}}^{\text{rel},1}(\vec{r}_{12})+
V_{\text{so}}^{\text{cm},1}(\vec{R}_{12}),
\end{eqnarray}
where
\begin{eqnarray}
\label{eq_vsorel}
V_{\text{so}}^{\text{rel},1}(\vec{r}_{12})=
-\imath \frac{\hbar^2 k_{\text{so}}}{m}\times \nonumber \\
\Bigg[
\left(
\frac{\partial}{\partial y_{12}} + \imath \frac{\partial}{\partial x_{12}}
\right)
| \uparrow \rangle_1 \, _1\langle \downarrow |
+ \nonumber \\
\left(
\frac{\partial}{\partial y_{12}} - \imath \frac{\partial}{\partial x_{12}}
\right)
| \downarrow \rangle_1 \, _1\langle \uparrow |
\Bigg]
\end{eqnarray}
and
\begin{eqnarray}
\label{eq_vsocm}
V_{\text{so}}^{\text{cm},1}(\vec{R}_{12})=
-\imath \frac{\hbar^2 k_{\text{so}}}{2m}\times \nonumber \\
\Bigg[
\left(
\frac{\partial}{\partial Y_{12}} + \imath \frac{\partial}{\partial X_{12}}
\right)
| \uparrow \rangle_1 \, _1\langle \downarrow |
+ \nonumber \\
\left(
\frac{\partial}{\partial Y_{12}} - \imath \frac{\partial}{\partial X_{12}}
\right)
| \downarrow \rangle_1 \, _1\langle \uparrow |
\Bigg].
\end{eqnarray}
In what follows, we drop the subscript $1$ of $|\uparrow \rangle_1$
and $|\downarrow \rangle_1$ and use $m_{s}$ instead of $m_{s1}$
for notational convenience.

To begin with, we consider case 1a with
$a_{\uparrow}=a_{\downarrow}=a_{\text{aa}}$.
We assume $N_{\text{cm}}=M_{\text{cm}}=K_{\text{cm}}=l_{\text{rel}}=m_{\text{rel}}=0$
and write the unperturbed states as $\Psi_{q_{\text{rel}},m_s}^{(0)}$,
where  $\Psi_{q_{\text{rel}},m_s}^{(0)}=
\psi_{q_{\text{rel}},0,0}^{\text{rel}} \psi_{0,0,0}^{\text{cm}} 
| m_s=\pm 1/2 \rangle$.
Moreover, we assume that
$\Psi_{q_{\text{rel}},m_s}^{(0)}$ is not
degenerate with any of the other unperturbed eigenstates
with the same $M_J$ and $K_{\text{cm}}$ quantum numbers.
This is fulfilled for all $q_{\text{rel}}\leq0$ [see the lowest
solid line in Fig.~\ref{fig_busch}(b)].
For $q_{\text{rel}}>0$ [for $0<q_{\text{rel}}<1/2$, e.g., 
see the lowest dashed line on the positive $a_{\text{aa}}$
side in Fig.~\ref{fig_busch}(b)], however,
degeneracies exist for selected $q_{\text{rel}}$ values.
Degeneracies also exist for all $q_{\text{rel}}=n-1/2$ 
($1/a_{\text{aa}}$=0; $n=1,2,3,...$)
and all $q_{\text{rel}}=n$ ($a_{\text{aa}}=0$; $n=1,2,3,...$).
In these cases, the coupling to other states
can notably enhance the interplay between the
spin-orbit coupling term and the $s$-wave interaction
(see Sec.~\ref{sec_degenerate}).
To treat the effect of $V_{\text{so}}(\vec{r}_1)$
in first-order non-degenerate perturbation theory,
we need to evaluate the
matrix element 
$\langle \Psi_{q_{\text{rel}},m_s}^{(0)} | V_{\text{so}} | 
\Psi_{q_{\text{rel}},m_s}^{(0)} \rangle$.
Since the states with different $m_s$
do not couple,
the first-order perturbation shift vanishes.

The second-order non-degenerate perturbation theory
expression contains
terms proportional to
$|\langle \Psi_{q_{\text{rel}},m_s}^{(0)} 
|V_{\text{so}}^{\text{rel},1}+V_{\text{so}}^{\text{cm},1} | \Psi_{\text{exc}}^{(0)} \rangle |^2$,
where the two-particle
state $\Psi_{\text{exc}}^{(0)}$ has a different
energy than $\Psi_{q_{\text{rel}},m_s}^{(0)}$.
It can be readily seen that terms that contain both $V_{\text{so}}^{\text{rel},1}$ and 
$V_{\text{so}}^{\text{cm},1}$ vanish due to the selection rules.
Terms that contain two $V_{\text{so}}^{\text{cm},1}$'s yield energy shifts
independent of $a_{\text{aa}}$. 
We evaluate these shifts using the techniques
discussed
in Sec.~\ref{sec_weakall}.
To evaluate the second-order perturbation theory expression
that contains two $V_{\text{so}}^{\text{rel},1}$'s,
we make three observations.
First, the integral over the center of mass coordinates only gives a non-zero
contribution when the $N_{\text{cm}}'$, $M_{\text{cm}}'$ and $K_{\text{cm}}'$ quantum
numbers that label the center of mass piece
of $\Psi_{\text{exc}}^{(0)}$ are equal to $0, 0$ and $0$, respectively.
Second, the integral over the relative coordinates
is only non-zero for
$m_{\text{rel}}'=\pm 1$, where the plus and minus signs apply if we assume that the
first particle is in the $m_s=1/2$ and $m_s=-1/2$ 
state, respectively.
Last, to evaluate the integrals involved,
we expand $\psi_{q_{\text{rel}},0,0}^{\text{rel}}$
in terms of non-interacting harmonic oscillator states~\cite{busc98, daily12},
\begin{eqnarray}
\label{eq_expandni}
\psi^{\text{rel}}_{q_{\text{rel}},0,0}(\vec{r}_{12})|m_s \rangle=
\sum_{j=0}^{\infty} C_j^{q_{\text{rel}}} \psi_{j,0,0,m_s}^{\text{rel}}(\vec{r}_{12}),
\end{eqnarray}
where the $\psi_{j,0,0,m_s}^{\text{rel}}$ are a product of
the non-interacting harmonic oscillator states and the spin part
(these states correspond---as mentioned above---to $q_{\text{rel}}=0,1,\cdots$)
and where the $C_j^{q_{\text{rel}}}$
denote expansion coefficients whose functional form is given in Eq.~(B8)
of Ref.~\cite{daily12} (see also Ref.~\cite{busc98}).
Using the expansion given in Eq.~(\ref{eq_expandni}),
the non-vanishing matrix elements are
$\langle \psi_{j,l_{\text{rel}}\mp1,0,\pm1/2}^{\text{rel}} 
| V_{\text{so}}^{\text{rel},1} | \psi_{j,l_{\text{rel}},\pm1,\mp1/2}^{\text{rel}} \rangle$
and
$\langle \psi_{j\pm1,l_{\text{rel}}\mp1,0,\pm1/2}^{\text{rel}} 
| V_{\text{so}}^{\text{rel},1} | \psi_{j,l_{\text{rel}},\pm1,\mp1/2}^{\text{rel}} \rangle$.
The matrix elements 
involved in second- and fourth-order
perturbation theory read
\begin{eqnarray}
\langle \psi_{j,0,0,\pm1/2}^{\text{rel}} 
| V_{\text{so}}^{\text{rel},1} | \psi_{j,1,\pm1,\mp1/2}^{\text{rel}} \rangle=
\nonumber \\
-\sqrt{\frac{2j+3}{6}} k_{\text{so}}a_{\text{ho}} E_{\text{ho}}, 
\end{eqnarray}
\begin{eqnarray}
\langle \psi_{j+1,0,0,\pm1/2}^{\text{rel}} 
| V_{\text{so}}^{\text{rel},1} | \psi_{j,1,\pm1,\mp1/2}^{\text{rel}} \rangle=
\nonumber \\
-\sqrt{\frac{j+1}{3}} k_{\text{so}}a_{\text{ho}} E_{\text{ho}}, 
\end{eqnarray}
\begin{eqnarray}
\langle \psi_{j,2,0,\pm1/2}^{\text{rel}} 
| V_{\text{so}}^{\text{rel},1} | \psi_{j,1,\pm1,\mp1/2}^{\text{rel}} \rangle=
\nonumber \\
-\sqrt{\frac{2j+5}{30}} k_{\text{so}}a_{\text{ho}} E_{\text{ho}},
\end{eqnarray}
and
\begin{eqnarray}
\langle \psi_{j-1,2,0,\pm1/2}^{\text{rel}} 
| V_{\text{so}}^{\text{rel},1} | \psi_{j,1,\pm1,\mp1/2}^{\text{rel}} \rangle=
\nonumber \\
-\sqrt{\frac{j}{15}} k_{\text{so}}a_{\text{ho}} E_{\text{ho}}.
\end{eqnarray}

Using these expressions 
in the second-order perturbation theory treatment of $k_{\text{so}}$,
we find that the infinite sum can be performed analytically.
Surprisingly, we find that the sum that involves two 
$V_{\text{so}}^{\text{rel},1}$'s
reduces to an expression that is independent of
$q_{\text{rel}}$. This implies that the
single particle spin-orbit term is not coupled to the $s$-wave 
interactions at this order of perturbation theory.
Combining the contributions that contain two $V_{\text{so}}^{\text{rel},1}$'s and those 
that contain two $V_{\text{so}}^{\text{cm},1}$'s, we find
\begin{eqnarray}
\label{eq_shift_soca1}
\Delta E^{(\text{so},2)}_{M_J=\pm1/2}=-(k_{\text{so}}a_{\text{ho}})^2E_{\text{ho}}.
\end{eqnarray}
This result is consistent with what we found 
in Eqs.~(\ref{eq_shift2}) 
and (\ref{eq_shiftgs}).

\begin{figure}
\includegraphics[angle=0,width=65mm]{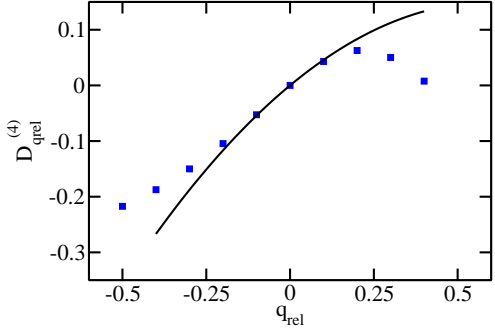}
\vspace*{0.2in}
\caption{(Color online)
Interplay between the 
$s$-wave
interaction and the spin-orbit coupling term
for the ground state for one
atom with and one atom without spin-orbit coupling
(case 1a; $\eta=1$).
The solid line and squares show the
quantity $D_{q_{\text{rel}}}^{(4)}$
that characterizes the fourth-order perturbation theory shift
as a function of $q_{\text{rel}}$.
The solid line shows 
the expansion around $q_{\text{rel}}=0$ [see Eq.~(\ref{eq_shiftso4pt})].
The squares show the full numerically determined values.
}
\label{fig_coupling}
\end{figure}

It can be shown that the third-order energy shift vanishes.
We find that the leading-order term that
reflects the interplay between the spin-orbit coupling term
and the $s$-wave interaction arises
at fourth-order perturbation theory,
\begin{eqnarray}
\label{eq_shiftso4}
\Delta E^{(\text{so},4)}_{M_J=\pm1/2}=
\left(\frac{1}{2}+D_{q_{\text{rel}}}^{(4)} \right) (k_{\text{so}}a_{\text{ho}} )^4E_{\text{ho}}.
\end{eqnarray}
The coefficient $D_{q_{\text{rel}}}^{(4)}$ 
depends on $q_{\text{rel}}$ and needs to be evaluated numerically.
Squares in Fig.~\ref{fig_coupling}
show the coefficient $D_{q_{\text{rel}}}^{(4)}$
as a function of $q_{\text{rel}}$.
When the $s$-wave scattering length is negative
($q_{\text{rel}}<0$), the interplay between the spin-orbit coupling term
and the $s$-wave interaction lowers the energy.
For $q_{\text{rel}}>0$ ($q_{\text{rel}}\ll1$), the interplay
leads to an increase of the energy.
Interestingly, for $q_{\text{rel}}\approx0.4$
(or $a_{\text{aa}}\approx2a_{\text{ho}}$),
$D_{q_{\text{rel}}}^{(4)}$ vanishes. For yet larger $q_{\text{rel}}$,
$D_{q_{\text{rel}}}^{(4)}$ becomes negative.
As $q_{\text{rel}}$ approaches $1/2$, the validity
regime of our perturbative expression is,
as discussed in more detail in Sec.~\ref{sec_degenerate},
small due to the presence of nearly degenerate states.
The non-degenerate perturbation theory
treatment breaks down when $q_{\text{rel}}=1/2$
and $N_{\text{cm}}=M_{\text{cm}}=K_{\text{cm}}=l_{\text{rel}}=m_{\text{rel}}=0$
(see the discussion in the second paragraph of this section),
i.e., when the two-body energy of the unperturbed
state equals $4\hbar \omega$.

In the weakly-interacting regime (small
$|a_{\text{aa}}|/a_{\text{ho}}$), an expansion 
around the non-interacting ground state, i.e., around $q_{\text{rel}}=0$, yields
\begin{eqnarray}
\label{eq_shiftso4pt}
D_{q_{\text{rel}}}^{(4)}=
\left[\frac{1}{4}-0.023(1)\frac{a_{\text{aa}}}{a_{\text{ho}}}+\cdots \right]
\frac{E_{\text{scatt}}}{E_{\text{ho}}},
\end{eqnarray}
where the coefficient of the 
$a_{\text{aa}}/a_{\text{ho}}$ term is
calculated numerically.
The first term in square brackets on the right hand side
of Eq.~(\ref{eq_shiftso4pt})
 agrees
with Eq.~(\ref{eq_shiftgs}) of Sec.~\ref{sec_pert_atomimp}.
The expansion
[the solid line in Fig.~\ref{fig_coupling}
shows Eq.~(\ref{eq_shiftso4pt})]
agrees well with the full expression for $|q_{\text{rel}}| \lesssim 0.1$.
For $q_{\text{rel}}=-1/2$, i.e., at unitarity, 
we find $D_{q_{\text{rel}}}^{(4)}=-0.216(1)$.

\begin{figure}
\vspace*{0.3in}
\includegraphics[angle=0,width=65mm]{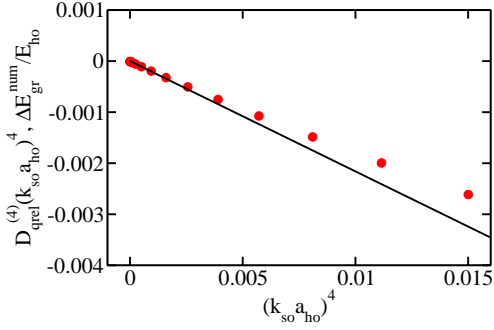}
\caption{(Color online)
Interplay between the 
$s$-wave
interaction and the spin-orbit coupling term
for the ground state for one
atom with and one atom without spin-orbit coupling
(case 1a with $1/a_{\text{aa}}=0$ and $\eta=1$).
The solid line
shows the quantity 
$D_{q_{\text{rel}}}^{(4)} (k_{\text{so}}a_{\text{ho}})^4$
for $q_{\text{rel}}=-1/2$
as a function of $(k_{\text{so}}a_{\text{ho}})^4$.
Circles show
the quantity 
$\Delta E_{\text{gr}}^{\text{num}}/E_{\text{ho}}$,
see Eq.~(\ref{eq_shift_numerics}).
}
\label{fig_compare_aiunit}
\end{figure}

Figure~\ref{fig_compare_aiunit} compares
the perturbative prediction (solid line) with the 
full numerical energy obtained
using the basis set expansion approach discussed  
in the Appendix for
$1/a_{\text{aa}}=0$ and $\eta=1$.
Circles show $\Delta E_{\text{gr}}^{\text{num}}$,
see Eq.~(\ref{eq_shift_numerics}),
as a function of $(k_{\text{so}}a_{\text{ho}})^4$.
The solid line in Fig.~\ref{fig_compare_aiunit} 
shows the scaled perturbative energy shift 
$D_{q_{\text{rel}}}^{(4)} (k_{\text{so}}a_{\text{ho}})^4$. 
The agreement is excellent
for $(k_{\text{so}}a_{\text{ho}})^4\lesssim0.004$ 
or $k_{\text{so}}a_{\text{ho}}\lesssim0.25$.

If we allow for different scattering lengths,
i.e., if we set $a_{\uparrow}=a_{\text{aa}}$ and 
$a_{\downarrow}=\eta a_{\text{aa}}$,
and assume $\eta\neq1$ (case 1b), then 
the two states
$\Psi_{q_{\text{aa}},1/2}^{(0)}$ and 
$\Psi_{q_{\eta \text{aa}},-1/2}^{(0)}$,
which have---as before---$N_{\text{cm}}
=M_{\text{cm}}=K_{\text{cm}}=l_{\text{rel}}=m_{\text{rel}}=0$,
have different energies.
Here, $q_{\text{aa}}$ and $q_{\eta \text{aa}}$
are the non-integer quantum numbers that
solve the transcendental equation
[Eq.~(\ref{eq_transcendental})] for
the states of interest
with $a_{\text{aa}}$ and $\eta a_{\text{aa}}$, respectively.
In what follows, 
we assume that $\Psi_{q_{\text{aa}}, 1/2}^{(0)}$ and 
$\Psi_{q_{\eta \text{aa}}, -1/2}^{(0)}$ are not
degenerate with any of the other unperturbed eigenstates
with the same $M_J$ quantum number.
In second-order perturbation theory,
the energy shifts, which are determined by
terms that contain two $V_{\text{so}}^{\text{cm},1}$'s,
depend on $q_{\text{aa}}$ and $q_{\eta \text{aa}}$.
Combining all second-order perturbation theory
contributions, we find that
the energy shift of the
unperturbed state $\Psi_{q_{\text{aa}},1/2}^{(0)}$ is given by
\begin{eqnarray}
\label{eq_shift_soca}
\Delta E^{(\text{so},2)}_{M_J=1/2}
=(-1+D_{q_{\text{aa}},q_{\eta \text{aa}}}^{(2)})
(k_{\text{so}}a_{\text{ho}})^2E_{\text{ho}},
\end{eqnarray}
where
\begin{eqnarray}
\label{eq_d2}
D_{q_{\text{aa}}, q_{\eta \text{aa}}}^{(2)}=
\frac{1}{2}+\frac{1}{2}
\sum_{q_{\text{rel}}}
\frac{(\sum_{j=0}^{\infty}
C_j^{q_{\text{aa}}}
C_j^{q_{\text{rel}}})^2}
{2q_{\text{aa}}-(2q_{\text{rel}}+1)}
\end{eqnarray}
and
$q_{\text{rel}}$ runs through all non-integer quantum numbers
that solve the transcendental equation
for $\eta a_{\text{aa}}$.
For $\eta=1$, $D_{q_{\text{aa}}, q_{\eta \text{aa}}}^{(2)}$
vanishes and Eq.~(\ref{eq_shift_soca})
reduces to Eq.~(\ref{eq_shift_soca1}).
In the weakly-interacting regime,
i.e., for small $|a_{\text{aa}}|/a_{\text{ho}}$ and
$|\eta a_{\text{aa}}|/a_{\text{ho}}$ 
($q_{\text{aa}}$ and $q_{\eta \text{aa}}$ near zero), 
Eq.~(\ref{eq_d2}) reduces to
\begin{eqnarray}
\label{eq_d2pt}
D_{q_{\text{aa}},q_{\eta \text{aa}}}^{(2)}=
\left(-\frac{1}{2}-\frac{2-\log4}{\sqrt{2\pi}}\frac{a_{\text{aa}}}{a_{\text{ho}}}
+\cdots \right) \times \nonumber \\
\left(1-\eta\right) \frac{E_{\text{scatt}}}{E_{\text{ho}}}.
\end{eqnarray}
The first term in large round brackets agrees with
the second term in square brackets in Eq.~(\ref{eq_shiftgssplit}).

To obtain the energy shift $\Delta E^{(\text{so},2)}_{M_J=-1/2}$ of the
unperturbed state $\Psi_{q_{\eta \text{aa}}, -1/2}^{(0)}$,
$D_{q_{\text{aa}},q_{\eta \text{aa}}}^{(2)}$ needs to be replaced
by $D_{q_{\eta \text{aa}},q_{\text{aa}}}^{(2)}$ in Eq.~(\ref{eq_shift_soca}),  
$q_{\text{aa}}$  needs to be replaced by $q_{\eta \text{aa}}$  in Eq.~(\ref{eq_d2}),
and $q_{\text{rel}}$ needs to run through all non-integer quantum numbers
that solve the transcendental equation
for $a_{\text{aa}}$.
In the weakly-interacting limit
($q_{\eta \text{aa}}$ and $q_{\text{aa}}$ near zero),
$D_{q_{\eta \text{aa}},q_{\text{aa}}}^{(2)}$ reduces to
Eq.~({\ref{eq_d2pt}) with $a_{\text{aa}}$ replaced by $\eta a_{\text{aa}}$
and $1-\eta$ replaced by $\eta-1$.
For $\eta\neq1$ (case 1b), the third-order
perturbation theory yields zero and the fourth-order
treatment is not pursued here.

\begin{figure}
\vspace*{0.3in}
\includegraphics[angle=0,width=65mm]{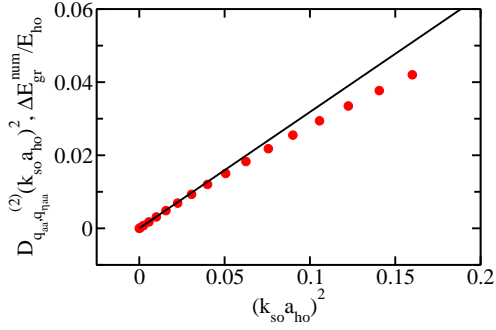}
\caption{(Color online)
Interplay between the 
$s$-wave
interaction and the spin-orbit coupling term
for the ground state for one
atom with and one atom without spin-orbit coupling
(case 1b with $1/a_{\text{aa}}=0$ and $\eta a_{\text{aa}}=0$).
The solid line
shows the quantity 
$D_{q_{\text{aa}},q_{\eta \text{aa}}}^{(2)} (k_{\text{so}}a_{\text{ho}})^2$
for $q_{\text{aa}}=-1/2$ and $q_{\eta \text{aa}}=0$
as a function of $(k_{\text{so}}a_{\text{ho}})^2$.
Circles show
the quantity 
$\Delta E_{\text{gr}}^{\text{num}}/E_{\text{ho}}$,
see Eq.~(\ref{eq_shift_numerics}).
}
\label{fig_compare_aiunitni}
\end{figure}

As an example, Fig.~\ref{fig_compare_aiunitni} compares
the perturbative prediction with the 
full numerical energy obtained
using the basis set expansion approach 
discussed in the Appendix for
$1/a_{\text{aa}}=0$ and $\eta a_{\text{aa}}=0$ (case 1b).
Circles show $\Delta E_{\text{gr}}^{\text{num}}/E_{\text{ho}}$,
see Eq.~(\ref{eq_shift_numerics}),
as a function of $(k_{\text{so}}a_{\text{ho}})^2$ while the
solid line
shows the scaled perturbative energy shift 
$D_{q_{\text{aa}},q_{\eta \text{aa}}}^{(2)} (k_{\text{so}}a_{\text{ho}})^2$. 
The agreement is excellent
for $(k_{\text{so}}a_{\text{ho}})^2\lesssim0.05$.
Figure~\ref{fig_compare_aiunitni} shows that
the interplay between the spin-orbit coupling term
and the $s$-wave interaction accounts for 
approximately 0.04$E_{\text{ho}}$ of the
energy for $(k_{\text{so}}a_{\text{ho}})^2=0.16$. This is a sizable effect
that should be measurable with present-day technology. 

To illustrate the behavior of the quantity 
$D_{q_{\text{aa}}, q_{\eta \text{aa}}}^{(2)}$, Eq.~(\ref{eq_d2}),
for other $q_{\text{aa}}$ and $q_{\eta \text{aa}}$ combinations,
squares in Figs.~\ref{fig_d2}(a)-\ref{fig_d2}(d) show 
$D_{q_{\text{aa}}, q_{\eta \text{aa}}}^{(2)}$ for 
$q_{\eta \text{aa}}=-1/2, -0.3, 0$ and $1/2$,
respectively, as a function of $q_{\text{aa}}$.
The solid line in Fig.~\ref{fig_d2}(c) shows the expansion
for small $|q_{\text{aa}}|$ and $|q_{\eta \text{aa}}|$
[see Eq.~(\ref{eq_d2pt})].
Interestingly, the expansion provides
a good description of the energy shift
over a fairly large range of $q_{\text{aa}}$ values.
For $q_{\text{aa}}<q_{\eta \text{aa}}$, the interplay 
between the spin-orbit coupling term and
$s$-wave interaction
leads to an increase of
the energy.
For $q_{\text{aa}}=q_{\eta \text{aa}}$
[$q_{\text{aa}}=-0.5, -0.3, 0$ and $0.5$
in Figs.~\ref{fig_d2}(a)-\ref{fig_d2}(d), respectively],
$D_{q_{\text{aa}}, q_{\eta \text{aa}}}^{(2)}$ vanishes.
For $q_{\eta \text{aa}}<q_{\text{aa}}<q_{\eta \text{aa}}+1/2$, the interplay
between the
spin-orbit coupling term and the $s$-wave interaction
leads to a decrease of the energy.
The behavior of $D_{q_{\text{aa}}, q_{\eta \text{aa}}}^{(2)}$
in the vicinity of the hashed regions is
discussed in the next section.

The key points of this section are:
\begin{itemize}
\item
For $a_{\uparrow}=a_{\downarrow}$ ($\eta=1$), 
the leading-order energy shift of the ground
state that reflects the interplay between the spin-orbit coupling
and the $s$-wave interaction is proportional to $(k_{\text{so}})^4$
for all scattering lengths.
\item
For $a_{\uparrow}\neq a_{\downarrow}$ ($\eta\neq1$), 
the leading-order energy shift of the ground
state that reflects the interplay between the spin-orbit coupling
and the $s$-wave interaction is, 
in general, proportional to $(k_{\text{so}})^2$
for all scattering lengths.
\end{itemize}

\begin{figure}
\vspace*{0.3in}
\includegraphics[angle=0,width=65mm]{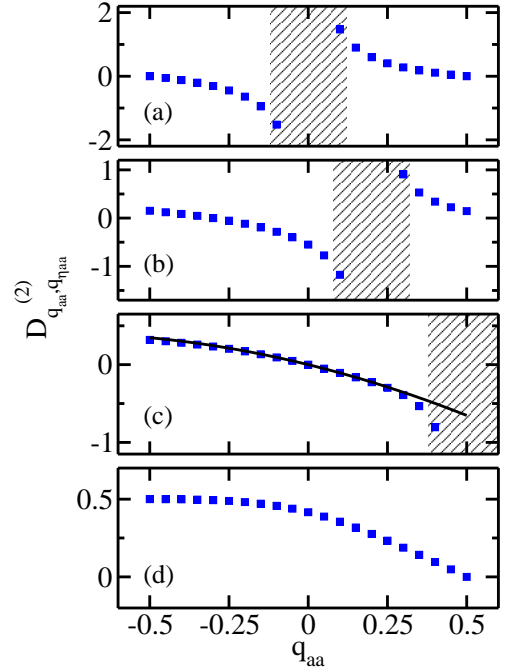}
\caption{(Color online)
Interplay between the 
$s$-wave
interaction and the spin-orbit coupling term for one
atom with and one atom without spin-orbit coupling
(case 1b).
The squares
show the numerically calculated quantity 
$D_{q_{\text{aa}},q_{\eta \text{aa}}}^{(2)}$
[see Eq.~(\ref{eq_d2})]
that characterizes the second-order perturbation theory
shift as a function of $q_{\text{aa}}$
for (a) $q_{\eta \text{aa}}=-1/2$,
(b) $q_{\eta \text{aa}}=-0.3$,
(c) $q_{\eta \text{aa}}=0$ and
(d) $q_{\eta \text{aa}}=1/2$.
The solid line in panel (c) shows the expansion for
small $|q_{\text{aa}}|$ and $|q_{\eta \text{aa}}|$
[see Eq.~(\ref{eq_d2pt})].
The hashed regions in panels (a)-(c)
show the parameter range where the non-degenerate
perturbation theory treatment breaks down.
}
\label{fig_d2}
\end{figure}

\subsection{Perturbative treatment of $V_{\text{so}}(\vec{r}_1)$:
Near-degenerate regime}
\label{sec_degenerate}
To understand the behavior of
$D_{q_{\text{aa}}, q_{\eta \text{aa}}}^{(2)}$
near the hashed regions in
Figs.~\ref{fig_d2}(a)-\ref{fig_d2}(c), it is
important to recall that the derivation
assumed that the states
$\Psi_{q_{\text{aa}}, 1/2}^{(0)}$ 
and $\Psi_{q_{\eta \text{aa}}, -1/2}^{(0)}$
are not degenerate with any other
unperturbed eigenstates with the same
$M_J$ quantum number.
To understand the implications,
we consider the situation where the unperturbed energy
equals $(2q_{\text{aa}}+3)\hbar\omega\leq4\hbar\omega$.
In this case, the $\Psi_{q_{\text{aa}}, 1/2}^{(0)}$ 
state with $N_{\text{cm}}=M_{\text{cm}}=K_{\text{cm}}=l_{\text{rel}}=m_{\text{rel}}=0$
($M_J=1/2$), referred to as state 1 in
the following, is degenerate with
the $M_J=1/2$ state with quantum 
numbers
$(q_{\eta \text{aa}}, l_{\text{rel}}, m_{\text{rel}}, N_{\text{cm}}, M_{\text{cm}}, K_{\text{cm}}, m_s)
=(q_{\text{aa}}-1/2, 0, 0, 0, 1, 0, -1/2)$,
referred to as state 2.
This degeneracy can be understood as follows.
Since the relative energy is equal to
$(2q_{\text{aa}}+3/2)\hbar \omega$ and 
$(2q_{\text{aa}}+1/2)\hbar \omega$ for states
1 and 2, respectively,
the unperturbed two-body energies are degenerate
if state 2 contains one ``extra" quantum of energy
in the center of mass degrees of freedom.
Putting this extra quantum in the $M_{\text{cm}}$
quantum number (as opposed to $K_{\text{cm}}$)
introduces a coupling between states 1 and 2 if the
spin-orbit coupling term is turned on.
In this case, the quantity
$D_{q_{\text{aa}}, q_{\eta \text{aa}}}^{(2)}$ does not provide a
faithful description of the energy spectrum for
$q_{\text{aa}}\approx q_{\eta \text{aa}}+1/2$,
i.e., for
$q_{\text{aa}}\approx0, 0.2$ and $1/2$
in Figs.~\ref{fig_d2}(a)-\ref{fig_d2}(c).
As discussed in the following, the coupling between states
1 and 2 leads to an enhancement of the interplay between
the spin-orbit coupling term and the $s$-wave interaction.

To determine the energy spectrum in the 
regime where states 1 and 2 have (near-)degenerate
energies, we employ first-order 
near-degenerate perturbation
theory~\cite{baym}. We define $\Delta$ through
$q_{\text{aa}}=q_{\eta \text{aa}}+1/2+\Delta$ and
assume $|\Delta| \ll1$.
We first diagonalize the Hamiltonian
$H_{\text{soc,a}}$
in the Hilbert space spanned by states 1 and 2. 
The diagonal matrix elements
are $(2q_{\text{aa}}+3)E_{\text{ho}}$ and 
$(2q_{\eta \text{aa}}+4)E_{\text{ho}}$ while
the off-diagonal elements are
$C_{q_{\text{aa}},q_{\eta \text{aa}}}^{(2)}
k_{\text{so}}a_{\text{ho}}E_{\text{ho}}/\sqrt{2}$,
where
\begin{eqnarray}
\label{eq_degc2}
C_{q_{\text{aa}},q_{\eta \text{aa}}}^{(2)}=\sum_{j=0}^{\infty}
C_j^{q_{\text{aa}}} C_{j}^{q_{\eta \text{aa}}}
\end{eqnarray}
and the $C_j$'s are defined through Eq.~(\ref{eq_expandni}).
The resulting first-order
energies are
\begin{eqnarray}
E/E_{\text{ho}}=2q_{\text{aa}}+3 - \Delta \pm \nonumber \\ 
\frac{1}{2} \sqrt{4\Delta^2+2(C_{q_{\text{aa}},q_{\eta \text{aa}}}^{(2)})^2
(k_{\text{so}}a_{\text{ho}})^2}.
\label{eq_degpt}
\end{eqnarray}
The second-order treatment then yields additional
shifts proportional to $(k_{\text{so}}a_{\text{ho}})^2$.

In the regime where the
energy difference between states 1 and 2
is much smaller than the coupling between
the two states ($|\Delta| \ll C_{q_{\text{aa}},q_{\eta \text{aa}}}^{(2)}
k_{\text{so}}a_{\text{ho}}/\sqrt{2}$),
we Taylor expand Eq.~(\ref{eq_degpt})
around $\sqrt{2}\Delta/(C_{q_{\text{aa}},q_{\eta \text{aa}}}^{(2)}
k_{\text{so}}a_{\text{ho}})=0$, 
\begin{eqnarray}
\label{eq_degpt_taylor1}
E/E_{\text{ho}}=2q_{\text{aa}}+3 - \Delta 
\pm
\frac{1}{\sqrt{2}} C_{q_{\text{aa}},q_{\eta \text{aa}}}^{(2)}
k_{\text{so}}a_{\text{ho}} \times \nonumber \\
\bigg[
1+\frac{\Delta^2}{(C_{q_{\text{aa}},q_{\eta \text{aa}}}^{(2)})^2
(k_{\text{so}}a_{\text{ho}})^2}+\cdots
\bigg].
\end{eqnarray}
For $\Delta=0$, Eq.~(\ref{eq_degpt_taylor1})
reduces to the result obtained using
degenerate perturbation theory.
Equation (\ref{eq_degpt_taylor1})
shows that the interplay between the $s$-wave
interaction and the spin-orbit coupling term
leads to an energy shift proportional to $k_{\text{so}}a_{\text{ho}}$.
In the regime
where the energy difference between states 1 and 2
is much greater than the coupling
($C_{q_{\text{aa}},q_{\eta \text{aa}}}^{(2)}
k_{\text{so}}a_{\text{ho}}/\sqrt{2} \ll |\Delta|$),
we Taylor expand Eq.~(\ref{eq_degpt})
around $C_{q_{\text{aa}},q_{\eta \text{aa}}}^{(2)}
k_{\text{so}}a_{\text{ho}}/(\sqrt{2} \Delta)=0$,
\begin{eqnarray}
\label{eq_degpt_taylor2}
E/E_{\text{ho}}=2q_{\text{aa}}+3-\Delta
\pm \Delta \times\nonumber \\
\bigg[
1+\frac{(C_{q_{\text{aa}},q_{\eta \text{aa}}}^{(2)})^2
(k_{\text{so}}a_{\text{ho}})^2}{4\Delta^2}
+\cdots
\bigg].
\end{eqnarray}
The eigenstates corresponding to
Eq.~(\ref{eq_degpt_taylor2}) are
approximately given by 
states 1 ($+$ sign) and 2 ($-$ sign),
respectively.
If we include the
second-order energy shift,
we recover our non-degenerate perturbation
theory results given in Eqs.~(\ref{eq_shift_soca}) and 
(\ref{eq_d2}).

\begin{figure}
\vspace*{0.3in}
\includegraphics[angle=0,width=65mm]{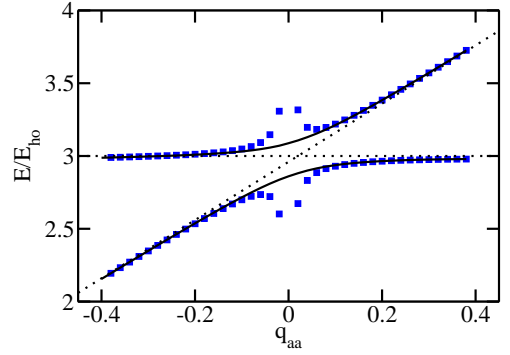}
\caption{(Color online)
Near-degenerate perturbation theory result
for one atom with and one atom without 
spin-orbit coupling (case 1b with $q_{\eta \text{aa}}=-1/2$
and $k_{\text{so}}a_{\text{ho}}=0.2$).
Dotted lines show the scaled energies
$(E_{\text{gr}}^{s-\text{wave}}+E_{\text{gr}}^{\text{so}})/E_{\text{ho}}-3$
for the two $M_J=1/2$ states (see text)
excluding the energy due to the interplay between
the spin-orbit coupling term and the $s$-wave interaction.
Solid lines show the energies
predicted by the near-degenerate perturbation theory
treatment up to second order.
For comparison,  
squares show the energies predicted by non-degenerate
perturbation theory [see Eq.~(\ref{eq_shift_soca}) for
the $M_{\text{cm}}=0$ state].}
\label{fig_deg}
\end{figure}

Figure~{\ref{fig_deg}} exemplarily illustrates
the results of the near-degenerate perturbation 
theory treatment 
for $k_{\text{so}}a_{\text{ho}}=0.2$, $q_{\eta \text{aa}}=-1/2$ 
and varying $q_{\text{aa}}$
[this corresponds to the hashed region
in Fig.~\ref{fig_d2}(a)].
The dotted lines show the scaled energies
$(E_{\text{gr}}^{s-\text{wave}}+E_{\text{gr}}^{\text{so}})/E_{\text{ho}}-3
=2q_{\text{aa}}+3-(k_{\text{so}}a_{\text{ho}})^2$ and $3$,
i.e., the energies of the system excluding the interplay between
the spin-orbit coupling term and the $s$-wave interaction.
The solid lines show the energies predicted
by the near-degenerate perturbation theory
treatment, including the first-order energies
[see Eq.~(\ref{eq_degpt})] 
and the second-order energy shifts 
[not given in Eq.~(\ref{eq_degpt})].
For $q_{\text{aa}}=0$, the first-order energies reduce to
$(3\pm \sqrt{1/\pi}k_{\text{so}}a_{\text{ho}})E_{\text{ho}}$.
The term proportional to $k_{\text{so}}a_{\text{ho}}$
reflects the interplay between
the spin-orbit coupling term and
the $s$-wave interaction. 
As can be
seen in Fig.~\ref{fig_deg},
the interplay turns the sharp crossing
(see dotted lines) into an avoided crossing (solid lines),
with the energy splitting governed by $k_{\text{so}}$.
The energy splitting
for $q_{\text{aa}}=0$ is roughly $0.2E_{\text{ho}}$.
This shift is much larger than the
energy shifts introduced by the interplay
between the
spin-orbit coupling term and
the $s$-wave interaction
for non-degenerate states.
This indicates that the interplay can,
for certain parameter combinations,
notably modify the energy spectrum even
for relatively small $|k_{\text{so}}|$.
For comparison,
the squares show the
second-order non-degenerate
perturbation theory energies.
The energy shift of 
state 1 is given in Eq.~(\ref{eq_shift_soca}) 
[see also Fig.~\ref{fig_d2}(a)] and the energy
shift of state 2 has been calculated following a similar
approach.

We note that there exist two 
other states with quantum numbers
$(q_{\eta \text{aa}}, l_{\text{rel}}, m_{\text{rel}}, N_{\text{cm}}, M_{\text{cm}}, K_{\text{cm}}, m_s)
=(q_{\text{aa}}-1/2, 0, 0, 0, -1, 0, -1/2)$
and 
$(q_{\eta \text{aa}}, l_{\text{rel}}, m_{\text{rel}}, N_{\text{cm}}, M_{\text{cm}}, K_{\text{cm}}, m_s)
=(q_{\text{aa}}-1/2, 0, 0, 0, 0, 1, -1/2)$
that have an energy of $(2q_{\text{aa}}+3)\hbar \omega$.
However, since these states have
$M_J=-3/2$ and $-1/2$, they do not couple
to the $M_J=1/2$ states discussed in
Eqs.~(\ref{eq_degc2})-(\ref{eq_degpt_taylor2}) and
Fig.~\ref{fig_deg}.
The $M_J=-3/2$ and $-1/2$ states
can be treated
using second-order non-degenerate perturbation theory.
In fact, the energy shift of the $M_J=-1/2$ state
is given in Eq.~(\ref{eq_shift_soca}). To get the energy shift
of the $M_J=-3/2$ state, the $-1$
in Eq.~(\ref{eq_shift_soca}) 
needs to be replaced by $-3/2$
and $D_{q_{\text{aa}}, q_{\eta \text{aa}}}^{(2)}$
needs to be multiplied by 2. The energy shifts 
of these two states are proportional to
$(k_{\text{so}}a_{\text{ho}})^2$ and their scaled energies
would be indistinguishable
from a horizontal
line on the scale of Fig.~\ref{fig_deg}.

The near-degenerate perturbation theory treatment
can be applied to other parameter combinations for
which degeneracies exist. As a second example,
we return to the system with $\eta=1$ (case 1a).
As stated earlier, Eq.~(\ref{eq_shiftso4}) does not
apply when $q_{\text{rel}}=1/2$
and $l_{\text{rel}}=m_{\text{rel}}=N_{\text{cm}}=M_{\text{cm}}=K_{\text{cm}}=0$,
i.e., when the two-body energy of the unperturbed
system equals $4\hbar \omega$.
In this case, the system supports six 
degenerate $M_J=1/2$ states. We find that these states do
not couple at first- and second-order perturbation
theory. However, the second-order treatment
yields energy shifts proportional to $-(k_{\text{so}}a_{\text{ho}})^2$
and $-(k_{\text{so}}a_{\text{ho}})^2/2$, thereby dividing the six
states into two smaller degenerate manifolds.
Treating these two manifolds separately, neither
of the states acquires a third-order shift. 
We notice, however,
that the states of these different manifolds
are, due to the shifts proportional to $-(k_{\text{so}}a_{\text{ho}})^2$,
degenerate at an energy less
than $4\hbar\omega$ (and a $q_{\text{rel}}$ value slightly
larger than $1/2$). Treating these new crossing
points, we find energy shifts proportional
to $k_{\text{so}}a_{\text{ho}}$ and 
avoided crossings governed by $(k_{\text{so}}a_{\text{ho}})^3$.

The discussion above shows that
the perturbative treatment of (avoided) crossings,
induced by the interplay between the
spin-orbit coupling term and the $s$-wave interaction,
requires great care. For the examples
investigated, we find that the interplay between
the spin-orbit coupling term and the $s$-wave interaction
gives rise to leading-order energy shifts proportional to odd
powers in $k_{\text{so}}a_{\text{ho}}$ in the vicinity of 
(avoided) crossings 
and to leading-order energy shifts proportional
to even powers in $k_{\text{so}}a_{\text{ho}}$ away from (avoided)
crossings.
We expect that the avoided crossings, introduced by the
interplay between the spin-orbit coupling term
and the $s$-wave interaction, have an appreciable
effect on the second-order virial coefficient and
related observables.

The key point of this section is:
\begin{itemize}
\item
The interplay between the spin-orbit coupling term 
and the $s$-wave interaction can,
if the energy levels of unperturbed states cross,
induce avoided crossings whose 
leading-order energy splitting
is proportional to $k_{\text{so}}$.
\end{itemize}

\subsection{Perturbative treatment of $V_{\text{so}}(\vec{r}_1)+V_{\text{so}}(\vec{r}_2)$:
Two particles with
spin-orbit coupling}
\label{sec_swavepert2}
This section considers two
particles with spin-orbit
coupling. As in Sec.~\ref{sec_swavepert1}, 
we rewrite the spin-orbit
coupling terms in terms of the relative and center
of mass coordinates,
\begin{eqnarray}
V_{\text{so}}(\vec{r}_1)
+V_{\text{so}}(\vec{r}_2)=
V_{\text{so}}^{\text{rel},1}(\vec{r}_{12})+
V_{\text{so}}^{\text{cm},1}(\vec{R}_{12}) \nonumber \\
-V_{\text{so}}^{\text{rel},2}(\vec{r}_{12})+
V_{\text{so}}^{\text{cm},2}(\vec{R}_{12}).
\end{eqnarray}
We assume $\omega_x=\omega_y=\omega_z$
and focus on the regime where center of mass excitations are absent
and where $l_{\text{rel}}=m_{\text{rel}}=0$.
As in Sec.~\ref{sec_swavepert1},  we account for the $s$-wave interaction
non-perturbatively. 

We start by considering case 2d, i.e., we consider the case
with $\zeta,\eta\neq1$ and $\zeta\neq\eta$,
and determine the perturbative shifts of the
states
$\Psi_{1/2, 1/2}^{(0)}=
\psi_{0,0,0}^{\text{cm}} \psi_{q_{\text{aa}},0,0}^{\text{rel}}
|\uparrow \rangle_1 |\uparrow \rangle_2$,
$\Psi_{-1/2, -1/2}^{(0)}=
\psi_{0,0,0}^{\text{cm}} \psi_{q_{\zeta \text{aa}},0,0}^{\text{rel}}
|\downarrow \rangle_1 |\downarrow \rangle_2$,
$\Psi_{1/2, -1/2}^{(0)}=
\psi_{0,0,0}^{\text{cm}} \psi_{q_{\eta \text{aa}},0,0}^{\text{rel}}
|\uparrow \rangle_1 |\downarrow \rangle_2$
and
$\Psi_{-1/2, 1/2}^{(0)}=
\psi_{0,0,0}^{\text{cm}} \psi_{q_{\eta \text{aa}},0,0}^{\text{rel}}
|\downarrow \rangle_1 \uparrow \rangle_2$
with $M_J=1, -1, 0$ and $0$, respectively.
Here $q_{\text{aa}}$, $q_{\zeta \text{aa}}$ and $q_{\eta \text{aa}}$ 
are obtained by solving the transcendental equation, 
Eq.~(\ref{eq_transcendental}), for $a_{\text{aa}}$,
$\zeta a_{\text{aa}}$ and $\eta a_{\text{aa}}$, respectively.

We assume that $\Psi_{1/2, 1/2}^{(0)}$
and $\Psi_{-1/2, -1/2}^{(0)}$ are not degenerate
with any other states with the same $M_J$.
Second-order
non-degenerate
perturbation theory then yields
\begin{eqnarray}
\label{eq_shift_socsoc1}
\Delta E^{(\text{so},2)}_{M_J=1(S)}=
2(-1+D_{q_{\text{aa}},q_{\eta \text{aa}}}^{(2)})
(k_{\text{so}}a_{\text{ho}})^2E_{\text{ho}}
\end{eqnarray}
for $M_J=1$.
The upper left arrow in Fig.~\ref{fig_socsoc}
schematically illustrates how the $M_J=1$ state
couples to the $M_J=0$ states.
The energy shift $\Delta E^{(\text{so},2)}_{M_J=-1(S)}$
is given by Eq.~(\ref{eq_shift_socsoc1})
with $q_{\text{aa}}$ replaced by $q_{\zeta \text{aa}}$.
The quantities $D_{q_{\text{aa}},q_{\eta \text{aa}}}^{(2)}$
and $D_{q_{\zeta \text{aa}},q_{\eta \text{aa}}}^{(2)}$
are defined in Eq.~(\ref{eq_d2})
and shown in Fig.~\ref{fig_d2}
for different $q$ combinations.
The states $\Psi_{1/2, -1/2}^{(0)}$ and 
$\Psi_{-1/2, 1/2}^{(0)}$
are degenerate.
Assuming no additional degeneracies with other $M_J=0$
states exist, degenerate perturbation theory yields
the second-order perturbation shifts
\begin{eqnarray}
\label{eq_shift_socsoc2}
\Delta E^{(\text{so},2)}_{M_J=0(S)}=2(-1+D_{q_{\eta \text{aa}},q_{\text{aa}}}^{(2)}+
\nonumber \\
D_{q_{\eta \text{aa}},q_{\zeta \text{aa}}}^{(2)})
(k_{\text{so}}a_{\text{ho}})^2E_{\text{ho}}
\end{eqnarray}
and 
\begin{eqnarray}
\label{eq_shift_socsoc3}
\Delta E^{(\text{so},2)}_{M_J=0(A)}=-2 (k_{\text{so}}a_{\text{ho}})^2E_{\text{ho}},
\end{eqnarray}
where
$D_{q_{\eta \text{aa}},q_{\text{aa}}}^{(2)}$ and 
$D_{q_{\eta \text{aa}},q_{\zeta \text{aa}}}^{(2)}$ are defined in
Eq.~(\ref{eq_d2}).
The eigenstates corresponding to Eqs.~(\ref{eq_shift_socsoc2})
and (\ref{eq_shift_socsoc3})
are respectively symmetric and anti-symmetric under the exchange of
particles 1 and 2.
The lower arrows in Fig.~\ref{fig_socsoc} schematically illustrate
the structure of Eq.~(\ref{eq_shift_socsoc2}).

\begin{figure}
\vspace*{0.1in}
\includegraphics[angle=0,width=70mm]{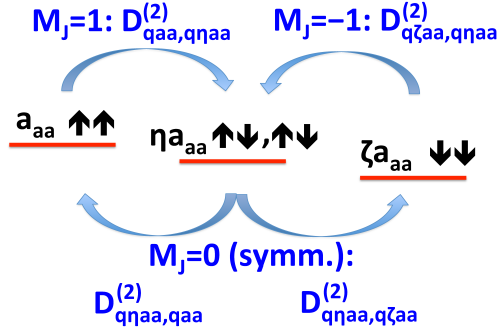}
\caption{(Color online)
Schematic illustration of Eqs.~(\ref{eq_shift_socsoc1})
and (\ref{eq_shift_socsoc2}). The horizontal
lines show the four unperturbed states under
consideration, labeled by their single particle spins
and scattering lengths (since the unperturbed
states $\Psi_{1/2, -1/2}^{(0)}$ and
$\Psi_{-1/2, 1/2}^{(0)}$ are degenerate, they are
represented by a single line);
the horizontal lines are vertically offset
to reflect the fact that they have different energies.
According to Eq.~(\ref{eq_shift_socsoc1}),
the coupling of the $M_J=1$ ($M_J=-1$) state
to the $M_J=0$ states is described by
$D_{q_{\text{aa}},q_{\eta \text{aa}}}^{(2)}$ ($D_{q_{\zeta \text{aa}},q_{\eta \text{aa}}}^{(2)}$).
According to Eq.~(\ref{eq_shift_socsoc2}),
the coupling of the symmetric $M_J=0$ state to the 
$M_J=1$ and $-1$ states is described
by $D_{q_{\eta \text{aa}},q_{\text{aa}}}^{(2)}$ and 
$D_{q_{\eta \text{aa}},q_{\zeta \text{aa}}}^{(2)}$, respectively.}
\label{fig_socsoc}
\end{figure}

In the weakly-interacting regime
(all $|q|$'s much smaller than 1), 
the $D^{(2)}$ coefficient
can be expanded [see Eq.~(\ref{eq_d2pt})].
The resulting energy shifts
proportional to $a_{\text{aa}}$ agree
with those derived in Sec.~\ref{sec_pert_bb}.
The treatment above breaks down 
when additional degeneracies exist. In this
case, near-degenerate perturbation theory
provides, in much the same way as
discussed in Sec.~\ref{sec_degenerate},
a reliable description of avoided crossings.

We find that Eqs.~(\ref{eq_shift_socsoc1})-(\ref{eq_shift_socsoc3})
hold in the limits that $\zeta$ or $\eta$ or both
go to 1.
For $\zeta=1$ and $\eta\neq1$ (case 2b),
the states $\Psi_{1/2, 1/2}^{(0)}$
and $\Psi_{-1/2, -1/2}^{(0)}$
are degenerate and have the same perturbation shift.
For $\zeta\neq1$ and $\eta=1$ (case 2c),
$D_{q_{\text{aa}}, q_{\eta \text{aa}}}^{(2)}$ vanishes.
The energy shift of the $M_J=1$ state contains
no term proportional to $(k_{\text{so}}a_{\text{ho}})^2$
while the energy shifts of the $M_J=-1$ and $M_J=0$ states 
with bosonic exchange symmetry contain shifts
proportional to $(k_{\text{so}}a_{\text{ho}})^2$.
In the limit that $\zeta=1$ and $\eta=1$ (case 2a),
$D_{q_{\text{aa}}, q_{\eta \text{aa}}}^{(2)}$,
$D_{q_{\zeta \text{aa}}, q_{\eta \text{aa}}}^{(2)}$,
$D_{q_{\eta \text{aa}}, q_{\text{aa}}}^{(2)}$ and
$D_{q_{\eta \text{aa}}, q_{\zeta \text{aa}}}^{(2)}$
vanish. In this case,
the interaction does not break the degeneracy
of the four unperturbed states and the energy shift
contains no term proportional to $(k_{\text{so}}a_{\text{ho}})^2$.

Figure~\ref{fig_compare_socsoc} 
compares the perturbative prediction (solid line)
with our numerical basis set expansion results (circles)
for case 2b. 
Figure~\ref{fig_compare_socsoc}(a) shows
the case where $1/a_{\text{aa}}=0$, $\zeta=1$ and $\eta a_{\text{aa}}=0$.
The lowest energy state is two-fold degenerate
($|M_J|=1$) and possesses bosonic exchange symmetry.
The leading-order energy shift that
reflects the interplay between the
spin-orbit coupling term and the $s$-wave interaction
is proportional to 
$(k_{\text{so}})^2$ [see Eq.~(\ref{eq_shift_socsoc1})].
Figure~\ref{fig_compare_socsoc}(b) shows
the case where $a_{\text{aa}}=0$, $\zeta=1$ and $1/(\eta a_{\text{aa}})=0$.
The lowest energy state possesses
fermionic exchange symmetry.
According to Eq.~(\ref{eq_shift_socsoc3}),
the interplay between 
the spin-orbit coupling term and
the $s$-wave interaction
does not give rise to an energy
shift proportional to $(k_{\text{so}})^2$.
This is confirmed by our numerical results (circles).

\begin{figure}
\vspace*{0.3in}
\includegraphics[angle=0,width=65mm]{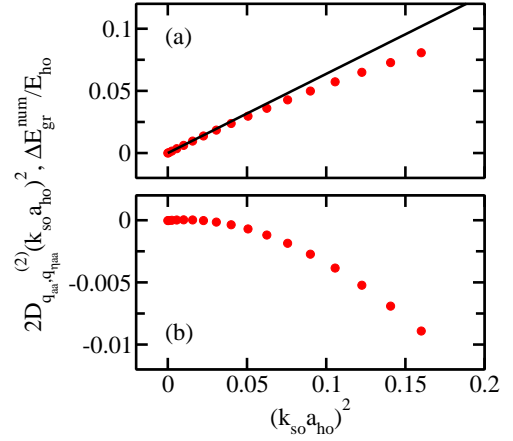}
\caption{(Color online)
Interplay between the 
$s$-wave
interaction and the spin-orbit coupling term
for the ground state manifold for two
atoms with spin-orbit coupling.
(a) The solid line 
shows the expression
$2D_{q_{\text{aa}},q_{\eta \text{aa}}}^{(2)}(k_{\text{so}}a_{\text{ho}})^2$,
see Eq.~(\ref{eq_shift_socsoc1}),
for the lowest energy state
as a function of
$(k_{\text{so}} a_{\text{ho}})^2$ for case 2b with
$1/a_{\text{aa}}=0$, $\zeta=1$ and $\eta a_{\text{aa}}=0$.
For comparison, the circles show the quantity
$\Delta E_{\text{gr}}^{\text{num}}/|E_{\text{scatt}}|$,
see Eq.~(\ref{eq_shift_numerics}).
(b) The circles show the quantity
$\Delta E_{\text{gr}}^{\text{num}}/|E_{\text{scatt}}|$ for
case 2b with $a_{\text{aa}}=0$,
$\zeta=1$
and $1/(\eta a_{\text{aa}})=0$.
The numerical data confirm the absence of
a term proportional to $(k_{\text{so}}a_{\text{ho}})^2$,
as predicted by Eq.~(\ref{eq_shift_socsoc3}).
}
\label{fig_compare_socsoc}
\end{figure}

The key points of this section are:
\begin{itemize}
\item
For two identical bosons, the energy shift proportional
to $(k_{\text{so}})^2$ is non-zero for the ground state
for all scattering lengths unless 
$a_{\uparrow\uparrow}=a_{\downarrow\downarrow}
=a_{\uparrow\downarrow}=a_{\downarrow\uparrow}$
$(\zeta=\eta=1)$
or, depending on the actual values of the scattering lengths,
$a_{\uparrow\uparrow}
=a_{\uparrow\downarrow}=a_{\downarrow\uparrow}$
$(\eta=1)$.
\item
For two identical fermions, the energy shift of the
ground state due to the interplay between
the spin-orbit coupling term and the $s$-wave interaction
does not contain a term proportional to $(k_{\text{so}})^2$
for any scattering lengths.  
\end{itemize}

\section{Conclusion}
\label{sec_conclusion}
For two point particles under external spherically symmetric
harmonic confinement with zero-range interaction, compact expressions
for the eigenenergies and eigenfunctions were obtained in 1998
by Busch and coworkers~\cite{busc98}.
These solutions (and the two- and one-dimensional analogs)
have played a crucial role in, to name a few examples,
analyzing few-atom experiments~\cite{esslinger05, 
esslinger06, jochim12},
guiding and benchmarking few-body calculations~\cite{blume07,
kestner07, vankolck07},
and interpreting the dynamics of many-body systems~\cite{blume03, bohn13}.
This paper determined portions of the energy spectrum of two
$s$-wave interacting atoms
under external spherically symmetric harmonic confinement with
spin-orbit coupling of Rashba type. The
spin-orbit coupling term introduces a new length scale as well
as new internal degrees of freedom or pseudo-spin states 
for the point particles
subject to the spin-orbit coupling.
Our calculations consider, building on the seminal work by Busch and coworkers~\cite{busc98},
two-atom systems with arbitrary $s$-wave scattering length
and small spin-orbit coupling strength.
We emphasize that the techniques developed
in this work can be adapted for treating non-spherical
traps, lower dimensional harmonic traps or
different spin-orbit coupling terms.
The treatment of anisotropic traps,
e.g., would utilize the analytical solutions
of Refs.~\cite{calarco05, calarco06}.

We obtained a large number of analytical results for
the small spin-orbit coupling strength regime.
Both the small and large scattering length regime were
considered. In the weakly-interacting regime, 
our results yield the leading-order mean-field shift.
For pure $s$-wave interactions the leading-order
mean-field shift of the trapped Bose gas is given
by $N(N-1)E_{\text{scatt}}$. Our calculations show how this
leading-order mean-field shift is modified
in the presence of a weak spin-orbit coupling term
of Rashba type.
At which order the leading interplay between the
spin-orbit coupling term and the $s$-wave interaction
arises depends strongly on whether or not
both particles feel the spin-orbit coupling
as well as on the actual values of the scattering lengths.
We discussed scenarios where
the leading-order interplay
between the spin-orbit 
coupling term and the $s$-wave interaction arises at order
$k_{\text{so}}$, $k_{\text{so}}^2$, $k_{\text{so}}^3$ and $k_{\text{so}}^4$.
A particularly strong interplay between the
spin-orbit coupling term and
the $s$-wave interaction was found in the vicinity of 
degeneracies, where the spin-orbit coupling term can
turn sharp crossings into avoided crossings.

Many of our perturbative results were 
validated by a numerical basis set
expansion approach for a wide range
of $s$-wave scattering lengths. 
Although most of our analysis was performed for the spin-orbit coupling
of Rashba type, the discussion in Sec.~\ref{sec_singleimpurity}
shows that at least some of our findings also
apply to systems with a spin-orbit coupling term of 
a different functional form.
For example, we found that, if only one of the particles
feels the spin-orbit coupling and 
$a_{\uparrow}=a_{\downarrow}=a_{\text{aa}}$,
the energy shift of the ground state
does not contain a term proportional to
$a_{\text{aa}} (k_{\text{so}})^2$. This result also holds for
anisotropic spin-orbit coupling of Rashba type 
and a spin-orbit coupling term that only involves the
$x$-component 
$p_{x}$ of the momentum.

Our analytical calculations employed a zero-range $s$-wave
model potential. To account for finite-range effects,
a momentum dependent term needs to be added.
For the weakly-interacting
trapped system, this yields an additional energy shift
proportional to $r_{\text{eff}}a_{\text{aa}}^2$, where $r_{\text{eff}}$
is the effective range~\cite{NJP2}. Our comparisons between
the numerical and perturbative results accounted for 
first- and higher-order effective range corrections non-perturbatively
by introducing the quantity $E_{\text{gr}}^{s-\text{wave}}$ in 
Eq.~(\ref{eq_shift_numerics}). 
In the weakly-interacting regime,
we find that the leading-order
interplay between the  spin-orbit coupling term and
the effective range
scales as $r_{\text{eff}}a_{\text{aa}}^2k_{\text{so}}^2$
(or higher order) for the ground state.
We estimate that this term,
for $|k_{\text{so}}|a_{\text{ho}}>|a_{\text{aa}}|/a_{\text{ho}}$, is
smaller than the terms that describe the interplay between 
the $s$-wave contact interaction and the spin-orbit 
coupling term considered in this paper.

It would be interesting to extend the perturbative and numerical calculations
presented 
in this paper to more than two particles. In pure $s$-wave systems,
effective three- and higher-body interactions have been 
shown to emerge~\cite{NJP1, NJP2}.
An intriguing question is how these effective few-body interactions
depend on the spin-orbit coupling term.
Another interesting question is how the thermodynamics of
Bose and Fermi gases with spin-orbit coupling differs from the 
thermodynamics of Bose and Fermi gases
without spin-orbit coupling.
A first answer to this question can be obtained by
looking at the virial equation of state
up to second order in the fugacity~\cite{ho04}. 
The virial equation of state
depends on the second-order virial coefficient, which
can be calculated if the complete energy spectrum 
of the trapped two particle system
is known~\cite{drummond09}.
Thus, a natural extension of the present work is to push the
two-particle calculations to higher energies and to larger 
spin-orbit coupling strengths. 
The large spin-orbit coupling regime has received a
great deal of attention recently.
In free space, the two-body binding energy has
been calculated and analytic expressions applicable
in weak and strong binding limits have been derived
\cite{magarill06, shenoy11, galitski12}.
It will be interesting to perform analogous calculations
for the trapped two-particle system with large $|k_{\text{so}}|a_{\text{ho}}$.

\section{Acknowledgements}
DB gratefully acknowledges J. Shertzer for providing an 
efficient iterative
generalized eigenvalue problem solver.
XYY and DB acknowledge
support by the National
Science Foundation (NSF) through Grant No.
PHY-1205443.
SG acknowledges support through
the Harvard Quantum Optics Center.
This work was additionally supported by the 
NSF through a grant for the Institute 
for Theoretical Atomic, Molecular and Optical Physics 
at Harvard University and
Smithsonian Astrophysical Observatory.

\appendix
\section{Basis set expansion approach}
\label{sec_numerics}
To determine the eigenenergies of the two-particle
system numerically, 
we expand the eigenstates in terms of basis functions that
contain
explicitly correlated Gaussians whose parameters are optimized 
semi-stochastically and solve the resulting 
generalized eigenvalue problem~\cite{cgbook, cgrmp}.
We first consider the situation where the first particle feels
the spin-orbit coupling while the second particle does not.
We write the eigenstate
$\Psi_{\text{soc,a}}(\vec{r}_1,\vec{r}_2)$
of the Hamiltonian $H_{\text{soc,a}}$
[see Eq.~(\ref{eq_ham_case1})] with 
$V_{\text{2b}}^{\sigma}(\vec{r}_{12})=V_{\text{g}}^{\sigma}(\vec{r}_{12})$
as
\begin{eqnarray}
\Psi_{\text{soc,a}}(\vec{r}_1,\vec{r}_2)=
\psi_{\uparrow}(\vec{r}_1,\vec{r}_2) | \uparrow \rangle_1
+
\psi_{\downarrow}(\vec{r}_1,\vec{r}_2) | \downarrow \rangle_1
\end{eqnarray}
and expand $\psi_{\uparrow}$ and $\psi_{\downarrow}$
in terms of geminals $g_j$~\cite{cgbook},
\begin{eqnarray}
\label{eq_wave_expand}
\psi_{\sigma}(\vec{r}_1,\vec{r}_2) = 
\sum_{j=1}^{N_b} 
c_j^{(\sigma)} g_j(\vec{R}, \underline{A}^{(j)}, \vec{s}^{(j)}),
\end{eqnarray}
where the $c_j^{(\sigma)}$ denote expansion coefficients and
$N_b$ denotes
the number of basis functions or geminals included in the
expansion.
The eigenstate of interest can be the ground state or an excited state.
The vector $\vec{R}$
collectively denotes the spatial degrees of freedom,
$\vec{R}=(\vec{r}_1,\vec{r}_2)$.

Each geminal $g_j$ is written in terms of a real and symmetric $2 \times 2$ 
matrix $\underline{A}^{(j)}$ and a six-component vector
$\vec{s}^{(j)}$,
$\vec{s}^{(j)} = (s_1^{(j)},\cdots,s_6^{(j)})$:
\begin{eqnarray}
g_j(\vec{R},\underline{A}^{(j)},\vec{s}^{(j)})=
\exp \left[
-\frac{1}{2} \vec{R}^T \underline{A}^{(j)} \vec{R} + (\vec{s}^{(j)})^T
\vec{R}
\right].
\end{eqnarray}
For concreteness, we write the 
argument of the exponential out explicitly;
we have
\begin{eqnarray}
(\vec{s}^{(j)})^T \vec{R} = s_1^{(j)} x_1 +s_2^{(j)} y_1 +\cdots
+
s_6^{(j)} z_2 
\end{eqnarray} 
and
\begin{eqnarray}
\vec{R}^T \underline{A}^{(j)} \vec{R} =
(A_{11}^{(j)} + A_{22}^{(j)} )
(\vec{r}_1^2 + \vec{r}_2^2)
+ \nonumber \\
2 A_{12}^{(j)} (x_1 x_2  + y_1 y_2 + z_1 z_2),
\end{eqnarray}
where $A_{kl}^{(j)}$ denotes the $kl$'s element of the
matrix $\underline{A}^{(j)}$.
The geminals $g_j$ have neither a definite orbital angular
momentum or projection quantum number nor a definite parity
and are thus suited to describe the eigenstates
of the two-particle system with spin-orbit coupling.
A key characteristic of the geminals is that the
Hamiltonian and overlap matrix elements reduce to
compact analytical expressions~\cite{cgbook}
if the atom-atom interaction is modeled by the Gaussian 
potential $V_{\text{g}}^{\sigma}$ [see Eq.~(\ref{eq_pot_gaussian})].

To construct the basis, we follow Ref.~\cite{debraj12}.
We start with just one basis function, i.e.,
we set $N_b=1$.
We calculate the $2 \times 2$ Hamiltonian and overlap matrices,
and diagonalize the resulting eigenvalue
problem.
In general, the Hamiltonian and overlap matrices have dimension
$(2 N_b) \times (2 N_b)$.
The factor of $2$ has its origin in the two internal degrees of freedom 
(pseudo-spin states) of the
first particle.
To add a new basis function,
we generate several thousand trial basis functions
semi-stochastically, i.e., we choose the
$A_{kl}^{(2)}$ and $s_k^{(2)}$
randomly from physically motivated preset ``parameter value
windows'', and select
the basis function that lowers the energy of the state
of interest the most.
This procedure is repeated till the basis set has reached the desired size,
i.e., till the energy of the state of interest is converged to the
desired accuracy.

The above approach generalizes readily to the situation where both particles 
feel the spin-orbit coupling [see Eq.~(\ref{eq_ham_case2}) for the 
Hamiltonian].
In this case, we write
\begin{eqnarray}
\Psi_{\text{soc,soc}}(\vec{r}_1,\vec{r}_2)= \nonumber \\
\psi_{\uparrow \uparrow}(\vec{r}_1,\vec{r}_2) 
| \uparrow \rangle_1 | \uparrow \rangle_2
+
\psi_{\uparrow \downarrow}(\vec{r}_1,\vec{r}_2) 
| \uparrow \rangle_1 | \downarrow \rangle_2
+
\nonumber \\
\psi_{\downarrow \uparrow}(\vec{r}_1,\vec{r}_2) 
| \downarrow \rangle_1 | \uparrow \rangle_2
+
\psi_{\downarrow \downarrow}(\vec{r}_1,\vec{r}_2) 
| \downarrow \rangle_1 | \downarrow \rangle_2
\end{eqnarray}
and expand the
$\psi_{\sigma \sigma'}(\vec{r}_1,\vec{r}_2)$
in terms of geminals [Eq.~(\ref{eq_wave_expand}) with
$\sigma$ replaced by $\sigma \sigma'$].
Since each particle has two internal degrees of freedom,
the overlap and Hamiltonian matrices that define the generalized eigenvalue
problem are $(4 N_b) \times (4 N_b)$-dimensional.

To validate our implementation, we performed
several checks:
{\em{(i)}}
We set the atom-atom potential to zero and determine the eigenenergies
for various $k_{\text{so}}$. We find that the ground state energy obtained by
the numerical basis set expansion approach agrees, within the
basis set extrapolation error, 
with the sum of the single-particle
energies (see Sec.~\ref{sec_singleimpurity} for the determination
of the single-particle energies). 
{\em{(ii)}}
We set $k_{\text{so}}=0$ and determine the eigenenergies for various depths
of the Gaussian model potential. In these calculations, we fix
$r_0$ at $r_0=0.02 a_{\text{ho}}$.
We find that the ground state energy obtained by the 
basis set expansion approach agrees, within the
basis set extrapolation error, with the energies obtained by a highly
accurate B-spline approach that separates the 
relative and center of mass degrees of freedom and takes advantage of the 
spherical symmetry of the system for $k_{\text{so}}=0$.
We find that the basis set expansion approach describes the 
two-particle systems with 
$a_{\text{ho}}/a_{\sigma} \lesssim 2 $ 
($a_{\text{ho}}/a_{\sigma \sigma'} \lesssim 2 $)
quite accurately.
In Secs.~\ref{sec_weakall} and \ref{sec_weaknotall},
we compare the energies obtained by the
basis set expansion approach with those obtained perturbatively
in the small
$|k_{\text{so}}|a_{\text{ho}}$ regime.
Our calculations reveal a rich interplay 
between the atom-atom interaction and the
spin-orbit coupling term.
The basis set expansion calculations reported in 
Secs.~\ref{sec_weakall} and \ref{sec_weaknotall} use $N_b \approx 200-400$.


\begin{thebibliography}{10}

\bibitem{rmp11}
J. Dalibard, F. Gerbier, G. Juzeli\=unas, and P. \"Ohberg,
Rev. Mod. Phys. {\bf{83}}, 1523 (2011).

\bibitem{spielmannat13}
V. Galitski	 and I. B. Spielman,
Nature {\bf{494}}, 49 (2013).

\bibitem{spielmanreview13}
N. Goldman, G. Juzeli\=unas, P. \"Ohberg, and I. B. Spielman,
Preprint at arXiv:1308.6533.

\bibitem{zhaireview}
H. Zhai, 
Int. J. Mod. Phys. B {\bf{26}}, 1230001 (2012).

\bibitem{spielman11}
Y.-J. Lin, K. Jim\'enez-Garc\'ia, and I. B. Spielman,
Nature {\bf{471}}, 83 (2011).

\bibitem{jzhang12}
P. Wang, Z.-Q. Yu, Z. Fu, J. Miao, L. Huang, S. Chai, H. Zhai,
and J. Zhang, Phys. Rev. Lett. {\bf{109}}, 095301 (2012).

\bibitem{jzhang13}
Z. Fu, L. Huang, Z. Meng, P. Wang, L. Zhang, S. Zhang, H. Zhai,
P. Zhang, and J. Zhang,
Preprint at arXiv:1306.4568.

\bibitem{spielman13}
R. A. Williams, M. C. Beeler, L. J. LeBlanc, K. Jim\'enez-Garc\'ia,
and I. B. Spielman, Phys. Rev. Lett. {\bf{111}}, 095301 (2013).

\bibitem{engels13}
C. Qu, C. Hamner, M. Gong, C. Zhang, and P. Engels,
Phys. Rev. A {\bf{88}}, 021604(R) (2013).

\bibitem{greene13}
A. Olson, S.-J. Wang, R. J. Niffenegger, C.-H. Li, C. H. Greene, and Y. P. Chen,
Preprint at arXiv:1310.1818.

\bibitem{zwierlein12}
L. W. Cheuk, A. T. Sommer, Z. Hadzibabic,
T. Yefsah, W. S. Bakr, and M. W. Zwierlein,
Phys. Rev. Lett. {\bf{109}}, 095302 (2012).

\bibitem{wu11}
C. Wu, I. Mondragon-Shem, and X.-F. Zhou, 
Chin. Phys. Lett. {\bf{28}}, 097102 (2011).

\bibitem{zhai10}
C. Wang, C. Gao, C. M. Jian, and H. Zhai, 
Phys. Rev. Lett. {\bf{105}}, 160403 (2010).

\bibitem{pu11}
L. Jiang, X.-J. Liu, H. Hu, and H. Pu,
Phys. Rev. A {\bf{84}}, 063618 (2011).

\bibitem{pu12}
B. Ramachandhran, B. Opanchuk, X.-J. Liu, H. Pu, P. D. Drummond, and H. Hu,
Phys. Rev. A {\bf{85}}, 023606 (2012).

\bibitem{galitski08}
T. D. Stanescu, B. Anderson, and V. Galitski,
Phys. Rev. A {\bf{78}}, 023616 (2008).

\bibitem{czhang11}
M. Gong, S. Tewari, and C. Zhang,
Phys. Rev. Lett. {\bf{107}}, 195303 (2011).

\bibitem{zhai11}
Z.-Q. Yu and H. Zhai,
Phys. Rev. Lett. {\bf{107}}, 195305 (2011).

\bibitem{sarang11}
S. Gopalakrishnan, A. Lamacraft, and P. M. Goldbart,
Phys. Rev. A {\bf{84}}, 061604(R) (2011).

\bibitem{baym12}
T. Ozawa and G. Baym,
Phys. Rev. A {\bf{85}}, 013612 (2012).

\bibitem{baym11}
T. Ozawa and G. Baym,
Phys. Rev. A {\bf{84}}, 043622 (2011).

\bibitem{williams11}
R. A. Williams, L. J. LeBlanc, K. Jim\'enez-Garc\'ia, 
M. C. Beeler, A. R. Perry, W. D. Phillips, and I. B. Spielman,
Science {\bf{335}}, 314 (2012).


\bibitem{pzhang12a}
P. Zhang, L. Zhang, and W. Zhang,
Phys. Rev. A {\bf{86}}, 042707 (2012).

\bibitem{pzhang12b}
P. Zhang, L. Zhang, and Y. Deng,
Phys. Rev. A {\bf{86}}, 053608 (2012).

\bibitem{pzhang13}
L. Zhang, Y. Deng, and P. Zhang,
Phys. Rev. A {\bf{87}}, 053626 (2013).

\bibitem{cui12}
X. Cui,
Phys. Rev. A {\bf{85}}, 022705 (2012).

\bibitem{gao13}
H. Duan, L. You, and B. Gao,
Phys. Rev. A {\bf{87}}, 052708 (2013).

\bibitem{jochim11}
F. Serwane, G. Z\"urn, T. Lompe, T. B. Ottenstein, A. N. Wenz,
and S. Jochim,
Science {\bf{332}}, 336 (2011).

\bibitem{bloch10}
S. Will, T. Best, U. Schneider, L. Hackerm\"uller, D.-S. L\"uhmann, and
I. Bloch,
Nature {\bf{465}}, 197 (2010).

\bibitem{2degexp}
M. Studer, G. Salis, K. Ensslin, D. C. Driscoll, and A. C. Gossard,
Phys. Rev. Lett. {\bf{103}},  027201 (2009).

\bibitem{2degtheory}
S. Chesi and G. F. Giuliani,
Phys. Rev. B {\bf{83}}, 235308 (2011).

\bibitem{qdexp}
Y. Kanai, R. S. Deacon,	 S. Takahashi, A. Oiwa,	 K. Yoshida,	K. Shibata, K. Hirakawa,	Y. Tokura, and S. Tarucha,
Nature Nanotechnology {\bf{6}}, 511 (2011).

\bibitem{governale02}
M. Governale,
Phys. Rev. Lett. {\bf{89}}, 206802 (2002).

\bibitem{pietilainen05}
T. Chakraborty and P. Pietil\"ainen,
Phys. Rev. B {\bf{71}}, 113305 (2005).

\bibitem{pi09}
E. Lipparini, M. Barranco, F. Malet, and M. Pi,
Phys. Rev. B {\bf{79}}, 115310 (2009).

\bibitem{reimann11}
A. Cavalli, F. Malet, J. C. Cremon, and S. M. Reimann,
Phys. Rev. B {\bf{84}}, 235117 (2011).

\bibitem{nanoexp}
S. Nadj-Perge, S. M. Frolov, E. P. A. M. Bakkers, and L. P. Kouwenhoven,
Nature {\bf{468}}, 1084 (2010).

\bibitem{busc98}
T. Busch, B.-G. Englert, K. Rz\c a\.zewski, and M. Wilkens,
Found. Phys. {\bf{28}}, 549 (1998).

\bibitem{rashba84}
Y. A. Bychkov and E. I. Rashba,
J. Phys. Chem. {\bf{17}}, 6039 (1984).

\bibitem{doucha87}
H. G. Reik, P. Lais, M. E. St\"utzle, and M. Doucha,
J. Phys. A {\bf{87}}, 6327 (1987).

\bibitem{analytic}
H. T\"ut\"unc\"uler, R. Ko\c{c}, and E. Ol\u{g}ar,
J. Phys. A {\bf{37}}, 11431 (2004).

\bibitem{chinrmp}
C. Chin, R. Grimm, P. Julienne, and E. Tiesinga,
Rev. Mod. Phys. {\bf{82}}, 1225 (2010).


\bibitem{hamnerthesis}
C. R. Hamner, Ph.D. thesis, Washington State University, 2014.


\bibitem{kramer}
H. A. Kramers, Proc. Amsterdam Acad. {\bf{33}}, 959 (1930).

\bibitem{degeneracytheorem}
M. J. Klein, Am. J. Phys. {\bf{20}}, 65 (1952).

\bibitem{zinner12}
O. V. Marchuov, A. G. Volosniev, D. V. Fedorov, A. S. Jensen,
and N. T. Zinner,
J. Phys. B {\bf{46}}, 134012 (2012).


\bibitem{NJP1}
P. R. Johnson, E. Tiesinga, J. V. Porto, and C. J. Williams,
New J. Phys. {\bf{11}}, 093022 (2009).

\bibitem{NJP2}
P. R. Johnson, D. Blume, X. Y. Yin, W. F. Flynn, and E. Tiesinga,
New J. Phys. {\bf{14}}, 053037 (2012). 

\bibitem{footnote}
As discussed in the text, the perturbation expression,
Eq.~(\ref{eq_shift1b}), which corresponds to an anti-symmetric 
state, is independent of $a_{\uparrow \uparrow}$ and
$a_{\downarrow \downarrow}$. The calculations based on the 
basis set expansion approach use a Gaussian potential
with the specified $a_{\sigma \sigma'}$ in the four scattering
channels. The fact that the numerical results are
well described by $a_{\uparrow \downarrow}$ alone
underlines the fact that the up-up and down-down channels are ``turned off" by the anti-symmetry of the wave function.  

\bibitem{daily12}
K. M. Daily, X. Y. Yin, and D. Blume,
Phys. Rev. A {\bf{85}}, 053614 (2012).

\bibitem{baym}
G. Baym.
{\em{Lectures on Quantum Mechanics}}.
Westview Press, Boulder (1969).

\bibitem{esslinger05}
H. Moritz, T. St\"oferle, K. G\"unter, M. K\"ohl,
and T. Esslinger,
Phys. Rev. Lett. {\bf{94}}, 210401 (2005).

\bibitem{esslinger06}
T. St\"oferle, H. Moritz, K. G\"unter, M. K\"ohl,
and T. Esslinger,
Phys. Rev. Lett. {\bf{96}}, 030401 (2006).

\bibitem{jochim12}
G. Z\"urn, F. Serwane, T. Lompe, A. N. Wenz, M. G. Ries, J. E. Bohn,
and S. Jochim,
Phys. Rev. Lett. {\bf{108}}, 075303 (2012).

\bibitem{blume07}
J. von Stecher, C. H. Greene, and D. Blume,
Phys. Rev. A {\bf{76}}, 053613 (2007).

\bibitem{kestner07}
J. P. Kestner and L.-M. Duan,
Phys. Rev. A {\bf{76}}, 033611 (2007).

\bibitem{vankolck07}
I. Stetcu, B. R. Barrett, U. van Kolck, and J. P. Vary,
Phys. Rev. A {\bf{76}}, 063613 (2007).

\bibitem{blume03}
B. Borca, D. Blume, and C. H. Greene,
New J. Phys. {\bf{5}}, 111 (2003).

\bibitem{bohn13}
A. G. Sykes, J. P. Corson, J. P. D'Incao, A. P. Koller, 
C. H. Greene, A. M. Rey, K. R. A. Hazzard, and J. L. Bohn,
Preprint at arXiv:1309.0828

\bibitem{calarco05}
Z. Idziaszek and T. Calarco, 
Phys. Rev. A {\bf{71}}, 050701(R) (2005).

\bibitem{calarco06}
Z. Idziaszek and T. Calarco, 
Phys. Rev. A {\bf{74}}, 022712 (2006).

\bibitem{ho04}
T.-L. Ho and E. J. Mueller,
Phy. Rev. Lett. {\bf{92}}, 160404 (2004).

\bibitem{drummond09}
X.-J. Liu, H. Hu, and P. D. Drummond,
Phys. Rev. Lett. {\bf{102}}, 160401 (2009).

\bibitem{magarill06}
A. V. Chaplik and L. I. Magarill,
Phys. Rev. Lett. {\bf{96}}, 126402 (2006).

\bibitem{shenoy11}
J. P. Vyasanakere and V. B. Shenoy,
Phys. Rev. B {\bf{83}}, 094515 (2011).

\bibitem{galitski12}
S. Takei, C.-H. Lin, B. M. Anderson, and V. Galitski,
Phys. Rev. A {\bf{85}}, 023626 (2012).

\bibitem{cgbook}
Y.~Suzuki and K.~Varga.
{\em{Stochastic Variational Approach to Quantum Mechanical Few-Body   
  Problems}}.
Springer Verlag, Berlin (1998).

\bibitem{cgrmp}
J. Mitroy, S. Bubin, W. Horiuchi, Y. Suzuki, L. Adamowicz,
W. Cencek, K. Szalewicz, J. Komasa, D. Blume, and K. Varga,
Rev. Mod. Phys. {\bf{85}}, 693 (2013).

\bibitem{debraj12}
D. Rakshit, K. M. Daily, and D. Blume,
Phys. Rev. A {\bf{85}}, 033634 (2012).
\end{thebibliography}
\end{document}